\documentclass[twoside]{article}
\usepackage{amsmath}			
\usepackage{amssymb}			
\usepackage{amsthm}			
\usepackage{a4wide}			
\usepackage{enumitem}			
\usepackage{slashed}			
\usepackage{feynmp}			
\usepackage{fancyhdr}          	 	
\usepackage[english]{babel}     	

\numberwithin{equation}{section}

\title{Real dimensional spaces in noncommutative geometry}
\author{Bas Jordans}
\date{\today}

\newtheorem{Thm}{Theorem}[section]
\newtheorem{Prop}[Thm]{Proposition}
\newtheorem{Lemma}[Thm]{Lemma}
\newtheorem{Cor}[Thm]{Corollary}

\theoremstyle{definition}
\newtheorem{Def}[Thm]{Definition}
\newtheorem{Exam}[Thm]{Example}
\newtheorem{Rem}[Thm]{Remark}
\newtheorem{Not}[Thm]{Notation}

\newenvironment{pf}{\vspace{0 pt plus 1pt minus 1pt}{\it Proof. }}{\hfill$\boxtimes$\vspace{6pt plus 1pt minus 2 pt}}

\newcommand{\Nat}{\mathbb{N}}
\newcommand{\Int}{\mathbb{Z}}

\newcommand{\Rea}{\mathbb{R}}
\newcommand{\Com}{\mathbb{C}}
\newcommand{\Acal}{\mathcal{A}}
\newcommand{\Bcal}{\mathcal{B}}

\newcommand{\Hcal}{\mathcal{H}}

\newcommand{\Kcal}{\mathcal{K}}
\newcommand{\Lcal}{\mathcal{L}}
\newcommand{\Mcal}{\mathcal{M}}
\newcommand{\Ncal}{\mathcal{N}}

\newcommand{\Pcal}{\mathcal{P}}

\newcommand{\Rcal}{\mathcal{R}}
\newcommand{\Scal}{\mathcal{S}}
\newcommand{\Tcal}{\mathcal{T}}

\newcommand{\Dz}{\slashed{D}_z}
\newcommand{\eps}{\varepsilon}
\newcommand{\ra}{\rightarrow}

\DeclareMathOperator{\Dom}{Dom}
\DeclareMathOperator{\ran}{ran}
\DeclareMathOperator{\Tr}{Tr}
\DeclareMathOperator{\im}{Im}
\DeclareMathOperator{\re}{Re}
\DeclareMathOperator{\sgn}{sgn}
\DeclareMathOperator{\clo}{clo}
\DeclareMathOperator{\Span}{span}
\DeclareMathOperator{\res}{res}

\newbox\ncintdbox \newbox\ncinttbox
\setbox0=\hbox{$-$}
\setbox2=\hbox{$\displaystyle\int$}
\setbox\ncintdbox=\hbox{\rlap{\hbox to \wd2{\kern-.1em\box2\relax\hfil}}\box0\kern.1em}
\setbox0=\hbox{$\vcenter{\hrule width 4pt}$}
\setbox2=\hbox{$\textstyle\int$}
\setbox\ncinttbox=\hbox{\rlap{\hbox to \wd2{\kern-.05em\box2\relax\hfil}}\box0\kern.1em}
\newcommand{\ncint}{\mathop{\mathchoice{\copy\ncintdbox}{\copy\ncinttbox}{\copy\ncinttbox}{\copy\ncinttbox}} \nolimits}

\fancypagestyle{fancyarticle}{
  \fancyhf{}                         
  \rhead[]{\rightmark}               
  \lhead[\leftmark]{}                
  \fancyhf[FLE,FRO]{\thepage}        
}

\fancypagestyle{plain}{ %
  \fancyhf{}                         
  
  \fancyhf[FLE,FRO]{\thepage}        
}

\begin{document}
\thispagestyle{plain}
\begin{center}
{\Large Real dimensional spaces in noncommutative geometry}

\bigskip
{\large Bas Jordans \footnote{Department of Mathematics, University of Oslo, P.O. Box 1053 Blindern, 0316 Oslo, Norway. \\ E-mail address: bpjordan@math.uio.no }}
\end{center}

\bigskip
\begin{abstract}
In this paper we will extend the product of spectral triples to a product of semifinite spectral triples. We will prove that finite summability and regularity are preserved under taking products. Connes and Marcolli constructed for each $z\in(0,\infty)$ a type ${\rm II}_\infty$-semifinite spectral triple which can be considered as a geometric space of dimension $z$. A small adaption of their construction yields a type ${\rm I}$-semifinite spectral triple. We will investigate the properties of these semifinite spectral triples. At the same time we will also avoid the need for an infra-red cutoff to compute the dimension spectrum. Using this collection of semifinite spectral triples and the product of semifinite spectral triples one can construct a mathematical tool for dimensional and zeta-function regularisation in quantum field theory.
\end{abstract}

\pagestyle{fancyarticle}
\section{Introduction}
In cite \cite{Hooft} Hooft and Veltman developed the method of dimensional regularisation to deal with divergent integrals in quantum field theory. The idea they had was to evaluated the corresponding integrals in $d-w$ dimensions for $w\in\Com$ instead of the original $d$ dimensions. This approach plays a key role in modern quantum field theory computations. It is therefore a natural question whether it is possible to mathematically construct geometric spaces which have dimension $z\in\Com$. As described by Connes and Marcolli in \cite{Connes2} this is indeed possible in the framework of noncommutative geometry. They found a spectral triple for which the `Dirac operator' $D$ satisfies the following generalization of the Gaussian integral in $z$ dimensions:
\[
\Tr\big(e^{-\lambda D^2}\big) = \Big(\frac{\pi}{\lambda}\Big)^{z/2}.
\]
Such spectral triples can be found in the generalisation of spectral triples to semifinite spectral triples, introduced by Benameur and Fack in \cite{Benameur}. \\
In the first section of this paper we will focus on these semifinite spectral triples, in particular we will consider the dimension spectrum \cite{Connes3} of semifinite spectral triples and we will deal with products of semifinite spectral triples. Combining this, one can obtain a semifinite spectral triple of `dimension' $n+z$ from the $z$-dimensional triple and a spin-manifold of dimension $n$. We will construct the product of semifinite spectral triples, this is a natural generalisation of the product of ordinary spectral triples and then we will show that the product preserves finite summability and regularity.
In the second part of this paper we will give the construction of the $z$-dimensional semifinite spectral triples. We will elaborate on why they can be considered as a generalisation of a geometric space of dimension $z$ and we will compare our construction with the definition of Connes and Marcolli. In particular we will pay attention to the `dimension spectrum', this is a subset of $\Com$ which is a generalisation of the dimension of manifolds to spectral triples. Connes and Moscovici \cite{Connes3} use the operator $|D|^{-1}$ in their definition of the dimension spectrum. This causes some problems, because the Dirac operator is not invertible in the triple we will consider. An alternative definition of the `dimension spectrum' uses the operator $(D^2+1)^{-1/2}$ which in this case is well-defined. However this definition has as a disadvantage that computations are a lot more complicated.
In the last part of this paper we will show that these semifinite spectral triples can indeed be used to define integrals in $z$ dimensions so that we obtain a little more concrete picture of dimensional regularisation. We will not build a full theory, but we will work out an example in details which illustrates the main ideas. Also we will show that it is possible to combine these $z$-dimensional spaces and the product of semifinite spectral triples to construct a tool for zeta-function regularisation.
An appendix is added in which the required results about traces on semifinite von Neumann algebras are stated.

\bigskip

{\bf Acknowledgements.}
This paper is based on the master's thesis I wrote at the Radboud University Nijmegen for the completion of my master in mathematics. I would like to thank Walter van Suijlekom for supervising me, the discussions we had and valuable comments he gave me during the work on my thesis and while writing down this paper.

\section{Semifinite noncommutative geometry}\label{semifinite}
The objective of section \ref{real_spaces} is to construct spectral triples which satisfy a specific requirement so that they can be considered to be $z$-dimensional for $z\in (0,\infty)$. The construction which is given makes use of semifinite traces and not the ordinary trace $\Tr$. This naturally leads to the notion of semifinite spectral triples. In this section we will derive some general results about these semifinite spectral triples, in particular we will be concerned with products of such triples.

\subsection{Semifinite spectral triples and their properties}
The difference between an ordinary spectral triple and a semifinite one is that we no longer require that the resolvent of the Dirac operator is compact, but we want it to be compact relative to a trace on a semifinite von Neumann algebra (c.f. definition \ref{tau_compact}).

\begin{Def}\label{semifinite_def}\cite{Benameur}
A {\it semifinite spectral triple} $(\Acal,\Hcal,D;\Ncal,\tau)$ consists of a Hilbert space $\Hcal$ a semifinite von Neumann algebra $\Ncal$ acting on $\Hcal$ with a faithful normal semifinite trace $\tau$, an involutive algebra $\Acal\subset\Ncal$ and a self-adjoint operator $D$ affiliated to $\Ncal$. Furthermore we require that for all $a\in\Acal$ the operator $[D,a]$ is densely defined and extends to a bounded operator on $\Hcal$ and that the operator $D$ is $\tau$-discrete. This operator $D$ is called the {\it Dirac operator} of the triple.\\
If in addition there exists a $\Int/2\Int$-grading $\gamma\in\Ncal$ such that $\gamma D = -D\gamma$ and $\gamma a=a\gamma$ for all $a\in\Acal$. Then the tuple $(\Acal,\Hcal,D;\Ncal,\tau,\gamma)$ is called an {\it even semifinite spectral triple}.
\end{Def}

The reason for the requirement $\gamma\in\Ncal$ is that we want that the trace $\tau(\gamma a)$ is defined for $a\in\Acal$. Almost all definitions of the classical case copy to the semifinite setting, in most cases we only have to deal with the substitution of $\Tr$ by a trace $\tau$.

\begin{Def}\label{summable}
Suppose $(\Acal,\Hcal,D;\Ncal,\tau)$ is a semifinite spectral triple. For $p>0$ we say that the triple is {\it $p$-$\tau$-summable} if $\tau((1+D^2)^{-p/2})<\infty$. The triple is {\it $\tau$-finitely summable} if it is $p$-$\tau$-summable for some $p>0$. The triple is {\it $p^+$-$\tau$-summable} if $\tau((1+D^2)^{-p/2+\eps})<\infty$ for all $\eps>0$. The triple is said to be {\it $\theta$-$\tau$-summable} if $\tau(e^{-tD^2})<\infty$ for any $t>0$.
\end{Def}

Note that since $D$ is self-adjoint, $\sigma(D^2)\subset[0,\infty)$ and thus $(D^2+1)^{-1}$ is well defined. These different notions of $\tau$-summability are related to one another in the following way.

\begin{Lemma}\label{notions_summability}
Suppose $(\Acal,\Hcal,D;\Ncal,\tau)$ is a semifinite spectral triple and $q>p>0$. If the triple is $p$-$\tau$-summable, then it is $p^+$-$\tau$-summable, $q$-summable and $\theta$-$\tau$-summable.
\end{Lemma}
\begin{pf}
Suppose $1<p<q$. Because $D$ is affiliated to $\Ncal$ and $(1+D^2)^{-1}$ is bounded, the operator $(1+D^2)^{-(q-p)/2}\in\Ncal$. Thus by \eqref{holder} the following inequality holds
\[
\tau\big((1+D^2)^{-q/2}\big) \leq \|(1+D^2)^{-(q-p)/2}\|\tau\big((1+D^2)^{-p/2}\big)<\infty.
\]
and thus $D$ is $q$-$\tau$--summable. If $\eps>0$, put $q:=p+2\eps$ from which $p^+$-$\tau$-summability follows. \\
We know that for $t>0$ and $\alpha>0$ fixed, the function $g_{t,\alpha}:[0,\infty)\ra\Rea$, $g_{t,\alpha}(x):= (1+x^2)^{\alpha/2} e^{-tx^2}$ is bounded, say by $C_{t,\alpha}$. Then
\begin{align*}
\tau\big(e^{-tD^2}\big) &= \tau\big(|g_{t,p}(D)|\,(1+D^2)^{-p/2}\big)\leq C_{t,p}\tau\big((1+D^2)^{-p/2}\big)<\infty,
\end{align*}
so the operator is $\theta$-$\tau$-summable and the last assertion follows.
\end{pf}

\begin{Def}\cite{Carey1}
For a semifinite spectral triple $(\Acal, \Hcal, D;\Ncal,\tau)$ we define for $a\in\Acal$ the operator $\delta(a):=[|D|,a]$, it is the unbounded derivation of $a$. We denote
\[
\Dom(\delta):=\{a\in B(\Hcal)\,:\, \delta(a) \textrm{ is bounded on }\Hcal \textrm{ and } a\Dom(|D|)\subset\Dom(|D|)\}.
\]
If for all $a\in\Acal$ the operators $a,[D,a]\in\Dom(\delta^k)$, we call the triple a $QC^k${\it-triple}, or $QC^k$ for short. If the triple is a $QC^k$-triple for all $k\geq 1$, we call it $QC^{\infty}$ or {\it regular}. An operator $a\in B(\Hcal)$ with $a\in\Dom(\delta^k)$ for all $k\in\Nat$ is called {\it smooth}.
\end{Def}

\begin{Def}
Suppose $(\Acal,\Hcal,D;\Ncal,\tau)$  is a regular $\tau$-finitely summable semifinite spectral triple. Let $\Bcal$ be the algebra generated by the elements $\delta^k(a)$ and $\delta^k([D,a])$ for $k\in\Nat$ and $a\in\Acal$. This triple is said to have {\it dimension spectrum} $Sd\subset\Com$ if for every $b\in\Bcal$ the zeta function $\zeta_{b}:z\mapsto \tau(b(1+D^2)^{-z/2})$ (for $\re(z)$ large), extends analytically to $\Com\setminus Sd$. If the set $Sd$ is discrete, then $(\Acal,\Hcal,D;\Ncal,\tau)$ is said to have a {\it discrete dimension spectrum}. If all zeta functions $\zeta_{b}$ have at most simple poles, the dimension spectrum is called {\it simple}.
\end{Def}

\begin{Rem}
There are two commonly used definitions of the dimension spectrum, namely the one introduced in \cite{Connes3} and the one stated above. In \cite{Connes3} the dimension spectrum is defined as the set of singularities of the the zeta functions
\[
z\mapsto \tau\big(b|D|^{-z}\big) \qquad (b\in\Bcal).
\]
However this definition gives problems if $D$ is not invertible, therefore we choose for the definition which uses $(D^2+1)^{-1/2}$ instead. The two different definitions are compatible with each other, see for instance \cite{Benameur}.
\end{Rem}

\subsection{Products of semifinite spectral triples}\label{semifinite_products}
In \cite{Connes6} the product of spectral triples is introduced. A detailed proof that this construction indeed yields a spectral triple can be found in e.g. \cite{Dabrowski}. We extend Connes' construction to a product of semifinite spectral triples. It is clear that we cannot expect this product to be a classical spectral triple, but we do obtain a semifinite spectral triple. This is the content of Theorem \ref{semifinite_product}. Such a theorem extends the classical product, because a classical spectral triple is a semifinite spectral triple with the type ${\rm I}$ von Neumann algebra $B(\Hcal)$ and trace $\Tr$. We start with the preparations for the proof of this theorem.

\begin{Lemma}\label{Takesaki_defIV1.2}{\normalfont \cite[\textsection IV.2]{Takesaki1}}
Let $\Hcal_1$ and $\Hcal_2$ be two Hilbert spaces. Suppose $\Hcal_2$ has an orthonormal basis $(e_i)_{i\in I}$. Denote $\Kcal_i$ for a copy of $\Hcal_1$. Then the map
\[
U:\bigoplus_{i\in I}\Kcal_i\ra \Hcal_1\otimes\Hcal_2,\qquad (x_i)_{i\in I}\mapsto\sum_{i\in I}x_i\otimes e_i
\]
is an isometry.
\end{Lemma}
\begin{pf}
Note that this sum is well-defined, because if $(x_i)_i\in \bigoplus_{i\in I}\Kcal_i$ at most countably many elements $x_i$ can be non-zero. Now clearly U is bijective and it preserves the norm. Hence it is an isometry.
\end{pf}

\begin{Prop}\label{product_semifinite_and_semifinite}
Let $\Mcal_1$ and $\Mcal_2$ be semifinite von Neumann algebras, then $\Mcal_1\otimes\Mcal_2$ is a semifinite von Neumann algebra.
\end{Prop}
\begin{pf}
See \cite[Thm. III.2.5.27]{Blackadar}.
\end{pf}

Since a von Neumann algebra is semifinite if and only if it admits a semifinite faithful normal trace \cite[Thm. V.2.15]{Takesaki1}, Proposition \ref{product_semifinite_and_semifinite} shows that given two semifinite von Neumann algebras $\Mcal_1$ and $\Mcal_2$ with semifinite faithful normal traces $\tau_1$ and $\tau_2$ there exists a semifinite normal trace on $\Mcal_1\otimes\Mcal_2$. The trace $\tau_1\otimes\tau_2$ indeed works.

\begin{Prop}\label{trace_on_product}
Suppose for $i=1,2$ $\Mcal_i$ is a semifinite von Neumann algebra with faithful semifinite normal trace $\tau_i$, then $\tau:=\tau_1\otimes\tau_2:(\Mcal_1\otimes\Mcal_2)_+\ra [0,\infty]$ is a faithful semifinite normal trace. In particular this trace factors, thus $\tau_1\otimes\tau_2(a_1\otimes a_2) = \tau_1(a_1)\tau_2(a_2)$.
\end{Prop}
\begin{pf}
From \cite[VIII.\textsection4]{Takesaki2} it follows that the map $\tau_1\otimes\tau_2$ is a faithful semifinite normal weight. So it remains to show that $\tau$ has the trace property, i.e. $\tau(aa^*)=\tau(a^*a)$ for all $a\in\Mcal_1\otimes\Mcal_2$. Let $a=\sum_n x_n\otimes y_n$, then
\[
aa^* = \Big(\sum_n x_n\otimes y_n\Big)\Big(\sum_n x_n\otimes y_n\Big)^* = \sum_n\sum_m x_nx_m^*\otimes y_ny_m^*.
\]
Since $\tau_i$ is a trace we have $\tau_i(ab)=\tau_i(ba)$ for all $a,b\in\Mcal_{i+}$ with $\tau_i(a),\tau_i(b)<\infty$. Therefore $\tau_i$ extends to a linear functional on $\Span\{a\in\Mcal_+\,:\,\tau_i(a)<\infty\}$. And if $a,b\in\Span\{a\in\Mcal_+\,:\,\tau_i(a)<\infty\}$ it holds hat $\tau_i(ab)=\tau_i(ba)$. Observe if $\tau(aa^*)=\infty$ then also $\tau(a^*a)=\infty$. So assume $\tau(aa^*)<\infty$. In that case $a=\sum_n x_n\otimes y_n$ with $|\tau_1(x_n)|<\infty$ and $|\tau_2(y_n)|<\infty$ for all $n$. Hence
\begin{align*}
\tau(aa^*)=\sum_n\sum_m \tau_1(x_nx_m^*)\tau_2(y_ny_m^*)
=\sum_n\sum_m \tau_1(x_m^*x_n)\tau_2(y_m^*y_n)
=\tau(a^*a).
\end{align*}
So indeed $\tau_1\otimes\tau_2$ is a trace.
\end{pf}

\begin{Lemma}\label{tensor_compact_operators}
In the notation of Lemma \ref{trace_on_product} suppose for $i=1,2$ the operators $K_i:\Hcal_i\ra\Hcal_i$ are $\tau_i$-compact. Then $K_1\otimes K_2:\Hcal_1\otimes\Hcal_2 \ra \Hcal_1\otimes\Hcal_2$ is a $\tau$-compact operator.
\end{Lemma}
\begin{pf}
Let $\eps>0$. Select for $i=1,2$ operators $R_i\in B(\Hcal_i)$ such that $\|K_i-R_i\|<\eps$, with the property that for the projection $P_i$ on the range of $R_i$ the trace $\tau_i(P_i)<\infty$. Then $P_1\otimes P_2$ is the projection on the range of $R_1\otimes R_2$. By the factorisation of $\tau$ (Lemma \ref{trace_on_product}) we have $\tau(P_1\otimes P_2)=\tau_1(P_1)\tau_2(P_2)<\infty$. By the cross-norm property of the norm on the tensor product
\begin{align*}
\|K_1\otimes K_2 - R_1\otimes R_2\|&\leq \|K_1\otimes K_2 - K_1\otimes R_2\| + \|K_1\otimes R_2 - R_1\otimes R_2\|\\
&\leq \|K_1\|\,\|K_2-R_2\| + \|K_1-R_1\|\,\|R_2\|\\
&\leq (\|K_1\|+\|K_2\|+\eps)\eps.
\end{align*}
\end{pf}

\begin{Thm}\label{semifinite_product}
Suppose $\Scal_1:=(\Acal_1,\Hcal_1,D_1;\Ncal_1,\tau_1,\gamma_1)$ is an even semifinite spectral triple and $\Scal_2:=(\Acal_2,\Hcal_2,D_2;\Ncal_2,\tau_2)$ is a semifinite spectral triple. Then
\[
\Scal:=(\Acal_1\otimes\Acal_2, \Hcal_1\otimes\Hcal_2,D_1\otimes1 + \gamma_1\otimes D_2; \Ncal_1\otimes\Ncal_2, \tau_1\otimes\tau_2)
\]
is a semifinite spectral triple. If in addition also $\Scal_2$ is even with a grading $\gamma_2$, then $\Scal$ is even with grading $\gamma_1\otimes\gamma_2$.
\end{Thm}

\begin{Def}
The triple $\Scal$ is called the {\it product} of the triples $\Scal_1$ and $\Scal_2$ and will be denoted by $\Scal_1\times\Scal_2$.
\end{Def}

\begin{Rem}
If we start with two even spectral triples $\Scal_1$ and $\Scal_2$, the triples $\Scal_1\times\Scal_2$ and $\Scal_2\times\Scal_1$ are related in the following way. The algebras, Hilbert spaces and von Neumann algebras of these two triples are isomorphic and the Dirac operators $D_1\otimes1+\gamma_1\otimes D_2$ and $D_1\otimes\gamma_2 + 1\otimes D_2$ are unitarily equivalent. Namely \cite{Vanhecke} for
\[
U:= \frac{1}{4}\,(1\otimes1 +\gamma_1\otimes1+1\otimes\gamma_2 - \gamma_1\otimes\gamma_2)
\]
it holds that
\[
U(D_1\otimes1+\gamma_1\otimes D_2)U^* = D_1\otimes\gamma_2+ 1\otimes D_2.
\]
\end{Rem}

We will now prove Theorem \ref{semifinite_product}.

\bigskip
\begin{pf}
The proof of this theorem is quite lengthy since we have to check several things. It is similar to the proof in \cite{Dabrowski} but it involves at lot more technical difficulties in particular the self-adjointness of the Dirac operator and compactness of its resolvent is difficult since we no longer have a basis of the Hilbert space consisting of eigenvectors of the Dirac operator.\\
For the ease of notion we introduce the following  objects
\[
\Acal:= \Acal_1\otimes\Acal_2, \qquad \Hcal:=\Hcal_1\otimes\Hcal_2, \qquad \Ncal:=\Ncal_1\otimes\Ncal_2, \qquad \tau:=\tau_1\otimes\tau_2, \qquad D:= D_1\otimes1+\gamma_1\otimes D_2.
\]
Where $\Dom(D):=\Dom(D_1)\otimes\Dom(D_2)$, the algebraic tensor product of vector spaces. If we have a second grading $\gamma_2$ on $\Scal_2$ we denote $\gamma:=\gamma_1\otimes\gamma_2$. Note that we are in the special situation that $D_1\otimes 1$ and $\gamma_1\otimes D_2$ anti-commute and therefore that $D^2=D_1^2\otimes1+1\otimes D_2^2$. \\
It is clear that $\Ncal$ is a von Neumann algebra acting on $\Hcal$. By Lemma \ref{product_semifinite_and_semifinite} $\Ncal$ is a semifinite von Neumann algebra and by Proposition \ref{trace_on_product} $\tau$ is a semifinite faithful normal trace. Concerning the algebra, the tensor product of two involutive algebras is again an involutive algebra. Since $\Acal_i\subset\Ncal_i$, the inclusion $\Acal\subset\Ncal$ is trivial. If we have a grading $\gamma_2$ on $\Ncal_2$, then obviously $\gamma=\gamma_1\otimes\gamma_2\in\Ncal_1\otimes\Ncal_2$. Before we can prove self-adjointness of the Dirac operator we will prove the next lemma.

\begin{Lemma}\label{Dom(D_1)tensor1}
The operators $D_1\otimes 1$ and $\gamma_1\otimes D_2$ with domains respectively $\Dom(D_1)\otimes\Hcal_2$ and $\Hcal_1\otimes\Dom(D_2)$ are self-adjoint.
\end{Lemma}
Note that $\Dom(D_1)\otimes \Hcal_2$ and $\Hcal_1\otimes\Dom(D_1)$ are algebraic tensor products of vector spaces and not tensor products of Hilbert spaces because $\Dom(D_i)$ are not Hilbert spaces.

\begin{pf}
As in Lemma \ref{Takesaki_defIV1.2} let $(e_i)_{i\in I}$ be an orthonormal basis of $\Hcal_2$ and let $U:\Hcal_1\otimes\Hcal_2\ra\bigoplus_{i\in I} \Kcal_i$ be the isometry $\sum_i x_i\otimes e_i\mapsto (x_i)_i$. Then
\[
U(D_1\otimes 1)U^*:\bigoplus_{i\in I} \Kcal_i\ra \bigoplus_{i\in I} \Kcal_i,\qquad (x_i)_i\mapsto (D_1 x_i)_i.
\]
Observe $U(\Dom(D_1)\otimes \Hcal_2)=\bigoplus_{i\in I} \Dom(D_1)_i$. Hence
\begin{align*}
\Dom(&(U(D_1\otimes 1)U^*)^*)\\
&= \Big\{(y_i)_i\in\bigoplus_{i\in I}\Kcal_i\,:\, \bigoplus_{i\in I}\Dom(D_1)_i\ra\Com;\; (x_i)_i\mapsto \sum_{i\in I}\langle D_1x_i,y_i\rangle \textrm{ is bounded}\Big\}.
\end{align*}
Suppose $(y_i)_i\in\bigoplus_{i\in I}\Kcal_i$ and there exists an $i_0\in I$ such that $y_{i_0}\notin\Dom(D_1)$. Then by self-adjointness of $D_1$ the map
\begin{equation}\label{eq-Dom(D_1)tensor_1}
\Dom(D_1)\ra\Com;\; x\mapsto \langle D_1x,y_{i_0}\rangle
\end{equation}
is unbounded. For an element $(x_i)_i$ with $x_i=0$ if $i\neq i_0$, we have $\sum_{i\in I}\langle D_1x_i,y_i\rangle = \langle D_1x_{i_0}, y_{i_0}\rangle$. Hence \eqref{eq-Dom(D_1)tensor_1} shows that for $y\notin\bigoplus_{i\in I}\Dom(D_1)_i$ the map
\[
\bigoplus_{i\in I}\Dom(D_1)_i\ra\Com;\; (x_i)_i\mapsto \sum_{i\in I}\langle D_1x_i,y_i\rangle
\]
is unbounded. Thus $\Dom((U(D_1\otimes 1)U^*)^*)\subset \bigoplus_{i\in I}\Dom(D_1)_i$. So $\Dom((D_1\otimes 1)^*)\subset\Dom(D_1)\otimes \Hcal_2$. For the converse inclusion let $y\in \Dom(D_1)\otimes\Hcal_2$, say $y=\sum_{n=1}^N y_1^{(n)}\otimes y_2^{(n)}$. Then for $x=\sum_{m=1}^M x_1^{(m)}\otimes x_2^{(m)}\in\Dom(D_1)\otimes\Hcal_2$ we have by self-adjointness of $D_1$
\begin{align}
\langle (D_1\otimes 1) x,y\rangle &= \sum_{n=1}^N\sum_{m=1}^M \langle D_1 x_1^{(m)},y_1^{(n)}\rangle\langle x_2^{(m)},y_2^{(n)}\rangle\notag\\
&= \sum_{n=1}^N\sum_{m=1}^M \langle x_1^{(m)},D_1y_1^{(n)}\rangle\langle x_2^{(m)},y_2^{(n)}\rangle\notag\\
&= \langle x,(D_1\otimes 1) y\rangle. \label{eq-Dom(D_1)tensor_2}
\end{align}
So $\Dom(D_1)\otimes \Hcal_2\ra\Com$, $x\mapsto \langle x,(D_1\otimes 1) y\rangle$, is bounded by $\|(D_1\otimes 1) y\|$. Hence $\Dom((D_1\otimes 1)^*)=\Dom(D_1)\otimes \Hcal_2$.\\
Similarly we have $\Dom((1\otimes D_2)^*)=\Hcal_1\otimes \Dom(D_2)$. The map $(\gamma_1\otimes1)^*=\gamma_1^*\otimes 1=\gamma_1\otimes1$ is bounded and maps $\Hcal_1\otimes\Dom(D_2)$ in itself, thus
\[
\Dom((\gamma_1\otimes D_2)^*)=\Dom((\gamma_1\otimes 1)^*(1\otimes D_2)^*)=\Hcal_1\otimes \Dom(D_2).
\]
The computation in \eqref{eq-Dom(D_1)tensor_2} shows $D_1\otimes1$ is symmetric, and we have $\Dom(D_1\otimes1)=\Dom(D_1)\otimes\Hcal_2=\Dom((D_1\otimes1)^*)$. So $D_1\otimes 1$ is self-adjoint on the domain $\Dom(D_1)\otimes \Hcal_2$. A similar argument applies to $\gamma_1\otimes D_2=(\gamma_1\otimes1)(1\otimes D_2)$.
\end{pf}

Now we are able to prove that $D$ is self-adjoint on the domain $\Dom(D_1)\otimes \Dom(D_2)$. The idea is to apply Nelson's theorem (c.f. Proposition \ref{Nelson}) to the operator $D$, from which we will obtain that $D$ is essentially self-adjoint and then we will show that $D$ is closed.\\
From Proposition \ref{Nelson} and the fact that $D_1$ and $D_2$ are self-adjoint we obtain that $\Dom^b(D_i)\subset \Hcal_i$ dense for $i=1,2$. Let $x\in\Dom^b(D_1)$ and $y\in\Dom^b(D_2)$. We will show that $x\otimes y\in\Dom^b(D_1\otimes 1+\gamma_1\otimes D_2)$. Select $C>0$ such that for all $n$
\[
\|D_1^nx\|\leq C^n \|x\|; \qquad \|D_2^ny\|\leq C^n \|y\|.
\]
Observe that $(D_1\otimes 1 + \gamma_1\otimes D_2)^2 = D_1^2\otimes 1 + 1\otimes D_2^2$ and that the operators $D_1^2\otimes1$  and $1\otimes D_2^2$ commute. Hence
\begin{align*}
\|D^{2n}(x\otimes y)\|&= \Big\|(D_1^2\otimes 1 + 1\otimes D_2^2)^n(x\otimes y)\Big\| \\
&= \Big\|\sum_{j=0}^n \binom{n}{j} D_1^{2j}x\otimes D_2^{2(n-j)}y\Big\|\\
&\leq \sum_{j=0}^n \binom{n}{j} \|D_1^{2j}x\| \|D_2^{2(n-j)}y\| \\
&\leq \sum_{j=0}^n \binom{n}{j} C^{2j}\|x\| C^{2(n-j)}\|y\| \\
&= (2C^2)^n \|x\otimes y\|\leq (2C)^{2n}\|x\otimes y\|.
\end{align*}
Similarly using $D^{2n+1} = (D_1\otimes 1 + \gamma_1\otimes D_2)\sum_{j=0}^n \binom{n}{j} D_1^{2j}\otimes D_2^{2(n-j)}$ gives
\[
\|D^{2n+1}(x\otimes y)\| \leq (2C)^{2n+1}\|x\otimes y\|.
\]
Hence $x\otimes y\in\Dom^b(D)$. Since $\Dom^b(D)$ is a linear subspace and $\Dom^b(D_i)\subset\Hcal_i$ dense, we have $\Dom^b(D_1)\otimes\Dom^b(D_2)\subset \Dom^b(D)\subset\Hcal_1\otimes\Hcal_2$ dense. But since $\Dom^b(D)\subset\Dom^a(D)$ then clearly $\Dom^a(D)\subset\Hcal$ dense. It is also clear that $D$ is symmetric, so $D$ is essentially self-adjoint. \\
It remains to show that $D$ is closed on $\Dom(D_1)\otimes\Dom(D_2)$. To that end, let $(x_n)_n$ be a sequence in $\Dom(D_1)\otimes\Dom(D_2)$ converging to $x\in\Hcal_1\otimes \Hcal_2$ such that $(Dx_n)_n$ is a Cauchy sequence in $\Hcal_1\otimes\Hcal_2$. For the moment fix $m,n$ and write $y=x_n-x_m$. Then $(D_1\otimes 1)y\in\Hcal_1\otimes \Dom(D_2)$ and $(\gamma_1\otimes D_2)y\in\Dom(D_1)\otimes\Hcal_2$. For ease of notation write $S:=\gamma_1\otimes D_2$ and $T:=D_1\otimes 1$. Then $STy=-TSy$ and thus
\begin{equation}\label{eq-Dirac_sa1}
\langle Dy,Dy\rangle = \langle Ty, Ty\rangle + \langle Sy, Sy\rangle.
\end{equation}
By Lemma \ref{Dom(D_1)tensor1} the operators $D_1\otimes 1$ and $\gamma_1\otimes D_2$ are self-adjoint and hence closed on the domains $\Dom(D_1)\otimes \Hcal_2$ respectively $\Hcal_1\otimes\Dom(D_2)$. The fact that $(Dx_n)_n$ is Cauchy combined with \eqref{eq-Dirac_sa1} gives that $(Tx_n)_n$ and $(Sx_n)_n$ are Cauchy. By closedness of the operators $T$ and $S$ we have $x\in(\Dom(D_1)\otimes\Hcal_2) \cap(\Hcal_1\otimes\Dom(D_2)) = \Dom(D_1)\otimes\Dom(D_2)$. Also the sequences $(Tx_n)_n$ and $(Sx_n)_n$ both have a limit in $\Hcal$, say $z_1$ respectively $z_2$ and $Tx=z_1$ and $Sx=z_2$. Thus
\[
\lim_{n\ra\infty}Dx_n=\lim_{n\ra\infty} Tx_n+Sx_n=z_1+z_2=Tx+Sx=Dx
\]
and $D$ is closed.

\bigskip

To prove that the operator $D$ is affiliated to the von Neumann algebra $\Ncal$, observe that by self-adjointness of $D$ it is sufficient to show that $\{D\}'\supset\Ncal'$. By assumption $\{D_i\}'\supset\Ncal_i'$. Also $\gamma_1\in\Ncal_1$ hence $\{\gamma_1\}'\supset \Ncal_1'$. Combination with the commutation theorem for tensor products (c.f. \cite[Thm. III.4.5.8]{Blackadar}) gives the inclusions
\[
\{D\}' \supset \{D_1\otimes 1\}' \cap \{\gamma_1\otimes D_2\}' \supset \Ncal_1'\otimes\Ncal_2'= (\Ncal_1\otimes \Ncal_2)' = \Ncal'
\]
and hence $D$ is affiliated to $\Ncal$.\\
Suppose $a\in\Acal_1\otimes\Acal_2$ and assume $a=a_1\otimes a_2$. Then
\[
[D,a_1\otimes a_2]= [D_1,a_1]\otimes a_2 - a_1\gamma_1\otimes[D_2,a_2].
\]
Because $[D_1,a_1]$ and $[D_2,a_2]$ are densely defined and bounded and the operators $a_2$ and $a_1\gamma_1$ are bounded, the commutator $[D,a_1\otimes a_2]$ is densely defined and extends to a bounded operator. Now for an arbitrary element $a=\sum_{i=1}^n a_i\otimes b_1$ we can show boundedness of the commutant $[D,a]$ using linearity of $D_i$ and bilinearity of $[\cdot,\cdot]$. Hence for all $a\in\Acal$ the operator $[D,a]$ is densely defined and extends to a bounded operator on $\Hcal$.\\
Before we will prove $\tau$-discreteness of $D$ we will need the results proven in the following two lemmas.

\begin{Lemma}\label{resolvent_of_sum}
The identity
\begin{equation}\label{eq-semifinite_product1}
(D+i)^{-1}=i(\gamma_1\otimes D_2+i)^{-1}(D_1\otimes1+i)^{-1} - (\gamma_1\otimes D_2+i)^{-1}(D_1\otimes1+i)^{-1}(\gamma_1\otimes D_2)(D_1\otimes1)(D+i)^{-1}
\end{equation}
holds as operators on $\Hcal$. Moreover $(\gamma_1\otimes D_2)(D_1\otimes1)(D+i)^{-1}$ is bounded.
\end{Lemma}
\begin{pf}
First we make a formal manipulation of the symbols. Then we have to check that the operators involved indeed extend to the whole Hilbert space $\Hcal$. \\
Again write $S:=\gamma_1\otimes D_2$ and $T:=D_1\otimes 1$. We have $ST=-TS$ and hence
\begin{align*}
\eqref{eq-semifinite_product1} &\Leftrightarrow (S+T+i)^{-1} = i(S+i)^{-1}(T+i)^{-1} - (S+i)^{-1}(T+i)^{-1}ST(S+T+i)^{-1}\\
&\Leftrightarrow 1=i(S+i)^{-1}(T+i)^{-1}(S+T+i)-(S+i)^{-1}(T+i)^{-1}ST\\
&\Leftrightarrow (T+i)(S+i)=i(S+T+i)-ST\\
&\Leftrightarrow TS +iS+iT-1 = iS +i T-1 -ST.
\end{align*}
The operators $(D,\Dom(D_1)\otimes\Dom(D_2))$, $(D_1\otimes1,\Dom(D_1)\otimes\Hcal_2)$ and $(\gamma_1\otimes D_2,\Hcal_2\otimes\Dom(D_2))$ are self-adjoint, thus $-i\notin\sigma(D),\sigma(D_1\otimes 1), \sigma(\gamma_1\otimes D_2)$. And hence $(D+i)^{-1}$, $(D_1\otimes1+i)^{-1}$ and $(\gamma\otimes D_2+i)^{-1}$ are bounded operators on $\Hcal$. Furthermore
\[
\ran((D+i)^{-1}) = \Dom(D+i) = \Dom(D) = \Dom(D_1)\otimes\Dom(D_2).
\]
Since $\gamma_1$ anticommutes with $D_1$
\begin{align*}
\Dom((D_1\otimes 1)(\gamma_1\otimes D_2)) &= \Dom(\gamma_1\otimes D_2)\cap (\gamma_1\otimes D_2)^{-1}(\Dom(D_1\otimes 1))\\
&= \Hcal_1\otimes\Dom(D_2) \cap \Dom(D_1)\otimes\Dom(D_2)\\
&= \Dom(D_1)\otimes\Dom(D_2).
\end{align*}
Observe that $\Dom\big(((\gamma_1\otimes D_2)(D_1\otimes 1))^*\big)\supset \Dom(D_1)\otimes\Dom(D_2)$. Hence $((\gamma_1\otimes D_2)(D_1\otimes 1))^*$ is densely defined and thus $((\gamma_1\otimes D_2)(D_1\otimes 1),\Dom(D_1)\otimes\Dom(D_2))$ is closable. We have previously shown that $\Dom(D+i)=\Dom(D)=\Dom(D_1)\otimes\Dom(D_2)$ and $D$ is self-adjoint on this domain. Hence the operator $(D+i)^{-1}$ maps $\Hcal$ into $\Dom(D_1)\otimes\Dom(D_2)$ and is bounded. Thus $(\gamma_1\otimes D_2)(D_1\otimes1)(D+i)^{-1}$ is a closed operator defined on $\Hcal$ and hence by the closed graph theorem it is a bounded operator. And equality \eqref{eq-semifinite_product1} holds on $\Hcal$.
\end{pf}

\begin{Lemma}\label{product_compact}
The operator $(\gamma_1\otimes D_2+i)^{-1}(D_1\otimes1+i)^{-1}: \Hcal_1\otimes\Hcal_2\ra \Hcal_1\otimes\Hcal_2$ is $\tau$-compact.
\end{Lemma}
\begin{pf}
The grading $\gamma_1$ satisfies, by definition, $\gamma_1^2=1$. This grading therefore induces a direct sum decomposition $\Hcal_1=\Hcal_1^+\oplus\Hcal_1^-$, where $\Hcal_1^\pm$ is the eigenspace of the eigenvalue $\pm1$ of $\gamma_1$. Then also
\[
\Hcal_1\otimes\Hcal_2\cong (\Hcal_1^+\otimes\Hcal_2)\oplus(\Hcal_1^-\otimes\Hcal_2).
\]
Now
\begin{align*}
\gamma_2\otimes D_2+i&:\Hcal_1^+\otimes\Hcal_2\ra\Hcal_1^+\otimes\Hcal_2, \qquad x\otimes y\mapsto x\otimes(D_2y+iy)\\
\gamma_2\otimes D_2+i&:\Hcal_1^-\otimes\Hcal_2\ra\Hcal_1^-\otimes\Hcal_2, \qquad x\otimes y\mapsto -x\otimes D_2y +x\otimes iy = -x\otimes(Dy-iy).
\end{align*}
Then the inverse of $\gamma_1\otimes D_2 + i$ is given in the matrix representation with respect to $\Hcal_1=\Hcal_1^+\oplus\Hcal_1^-$ as
\begin{equation}\label{eq-product_compact1}
(\gamma_1\otimes D_2+ i)^{-1} = \begin{pmatrix} 1& 0\\0&0\end{pmatrix}\otimes (D_2+i)^{-1} + \begin{pmatrix} 0& 0\\0&-1\end{pmatrix}\otimes (D_2-i)^{-1}.
\end{equation}
Since $D_1$ anti-commutes with $\gamma_1$ and $D_1$ is self-adjoint, in this decomposition the operator $D_1$ can be written as
\[
D_1=\begin{pmatrix} 0& D_1^+\\D_1^-&0\end{pmatrix},
\]
where $D_1^{+*}=D_1^-$. One can check that $(D_1\otimes I+i)^{-1}=(D_1+i)^{-1}\otimes I$ is given by the matrix
\begin{equation}\label{eq-product_compact4}
\begin{pmatrix} -i(1+D_1^+D_1^-)^{-1} & (D_1^-+(D_1^+)^{-1})^{-1}\\ ((D_1^-)^{-1}+D_1^+)^{-1} & -i(D_1^-D_1^++1)^{-1}\end{pmatrix} \otimes I.
\end{equation}
Multiplication of \eqref{eq-product_compact1} and  \eqref{eq-product_compact4} gives that $(\gamma_1\otimes D_2+i)^{-1}(D_1\otimes1+i)^{-1}$ is represented as
\begin{align}\label{eq-product_compact3}
&\begin{pmatrix} -i(1+D_1^+D_1^-)^{-1} & (D_1^-+(D_1^+)^{-1})^{-1}\\ 0 & 0\end{pmatrix} \otimes (D_2+i)^{-1}\notag\\
&\qquad + \begin{pmatrix} 0 & 0\\ -((D_1^-)^{-1}+D_1^+)^{-1} & i(D_1^-D_1^++1)^{-1}\end{pmatrix} \otimes (D_2-i)^{-1}.
\end{align}
Because $(D_1+i)^{-1}$ is $\tau_1$-compact, there exists a sequence $(F_n)_n$ of $\tau_1$-finite rank operators with $\lim_n \|(D+i)^{-1}-F_n\| = 0$. Decompose $F_n$ as a matrix with respect to $\Hcal_1=\Hcal_1^+\oplus\Hcal_1^-$ by
\[
F_n=\begin{pmatrix} F_n^{11} & F_n^{12} \\ F_n^{21} & F_n^{22} \end{pmatrix}.
\]
Thus from the estimate
\begin{align*}
\Big\| \Big( &\begin{pmatrix} -i(1+D_1^+D_1^-)^{-1} & (D_1^-+(D_1^+)^{-1})^{-1}\\ 0 & 0\end{pmatrix} - \begin{pmatrix} F_n^{11} & F_n^{12} \\ 0 & 0 \end{pmatrix}\Big)\begin{pmatrix} x\\ y\end{pmatrix}\Big\|\\
&= \Big\| \begin{pmatrix} (-i(1+D_1^+D_1^-)^{-1} - F_n^{11})x + ((D_1^-+(D_1^+)^{-1})^{-1}-F_n^{21})y \\ 0 \end{pmatrix}\Big\| \\
&\leq \Big\|(F_n-(D_1-i)^{-1})\begin{pmatrix} x\\ y\end{pmatrix}\Big\|
\end{align*}
and a similar one for the lower entries of the matrices we obtain
\begin{align*}
\lim_{n\ra\infty} \Big\| \begin{pmatrix} -i(1+D_1^+D_1^-)^{-1} & (D_1^-+(D_1^+)^{-1})^{-1}\\ 0 & 0\end{pmatrix} - \begin{pmatrix} F_n^{11} & F_n^{12} \\ 0 & 0 \end{pmatrix} \Big\| = 0; \\
\lim_{n\ra\infty} \Big\| \begin{pmatrix} 0 & 0\\ -((D_1^-)^{-1}+D_1^+)^{-1} & i(D_1^-D_1^++1)^{-1}\end{pmatrix} - \begin{pmatrix} 0 & 0 \\ -F_n^{21} & -F_n^{22} \end{pmatrix}\Big\|=0.
\end{align*}
Since the range of each of the operators $F_n^{ij}$ is contained in the range of $F_n$ ($i,j=1,2$) the operators $F_n^{ij}$ are $\tau_1$ compact. Therefore the operators represented by
\[
\begin{pmatrix} -i(1+D_1^+D_1^-)^{-1} & (D_1^-+(D_1^+)^{-1})^{-1}\\ 0 & 0\end{pmatrix} \qquad \begin{pmatrix} 0 & 0\\ -((D_1^-)^{-1}+D_1^+)^{-1} & i(D_1^-D_1^++1)^{-1}\end{pmatrix}
\]
are $\tau_1$ compact. By assumption $D_2$ is $\tau_2$-discrete, thus $(D_2+i)^{-1}$ and $(D_2-i)^{-1}$ are $\tau_2$-compact. Using Lemma \ref{tensor_compact_operators} the operator given by \eqref{eq-product_compact3} is $\tau$-compact, thus  $(D_1\otimes I+i)^{-1}=(D_1+i)^{-1}\otimes I$ is $\tau$-compact.
\end{pf}

Now it has become easy to prove compactness of the resolvent. Since $D$ is self-adjoint, $\sigma(D)\subset\Rea$. According to Theorem \ref{thm-tau-discrete} it is therefore sufficient to show that $(D+i)^{-1}$ is $\tau$-compact. Recall that the $\tau$-compact operators are an ideal in $B(\Hcal_1\otimes\Hcal_2)$. Combination of this fact with Lemmas \ref{resolvent_of_sum} and \ref{product_compact} imply that $(D+i)^{-1}$ is $\tau$-compact. This concludes the proof that $\Scal_1\otimes\Scal_2$ is a semifinite spectral triple.\\
In the case that we have a grading $\gamma_2$ on the second triple, for $\sum_k n_1^{(k)}\otimes n_2^{(k)}\in\Ncal$ we have
\[
(\gamma_1\otimes\gamma_2)\Big(\sum_k n_1^{(k)}\otimes n_2^{(k)}\Big) = \sum_k \gamma_1n_1^{(k)}\otimes \gamma_2n_2^{(k)} = \sum_k n_1^{(k)}\gamma_1\otimes n_2^{(k)}\gamma_2 = \Big(\sum_k n_1^{(k)}\otimes n_2^{(k)}\Big)(\gamma_1\otimes\gamma_2)
\]
and
\[
(\gamma_1\otimes\gamma_2)D = (\gamma_1D_1)\otimes \gamma_2 + \gamma_1^2\otimes(\gamma_2D_2) = -(D_1\gamma_1)\otimes\gamma_2 - \gamma_1^2\otimes (D_2\gamma_2) = -D(\gamma_1\otimes\gamma_2).
\]
Hence $\gamma$ is a grading on $\Scal$, thus the product of two even triples is again an even semifinite spectral triple.
\end{pf}

If one takes the product of two manifolds say of dimension $m$ and $n$, the product has dimension $m+n$. Therefore we might expect that the product of two finitely summable semifinite spectral triples is again finitely summable. This is indeed true.

\begin{Lemma}\label{summability_product}
Suppose for $i=1,2$ the tuples $(\Acal_1,\Hcal_i,D_i;\Ncal_i,\tau_i)$ are semifinite spectral triples and the first triple is even with grading $\gamma_1$. If the triples are $p_i$-$\tau_i$-summable, then the semifinite product spectral triple is $(p_1+p_2)$-$\tau$-summable. If both spectral triples are $\theta$-$\tau_i$-summable, then the product spectral triple is $\theta$-$\tau$-summable.
\end{Lemma}
\begin{pf}
To prove the first statement, suppose that both triples are $p_i$-$\tau_i$-summable. Note that it holds that $1+D_1^2\otimes 1\leq 1+D_1^2\otimes1 +1\otimes D_2^2$ and similarly $1+1\otimes D_2^2\leq 1+D_1^2\otimes1 +1\otimes D_2^2$, thus
\[
(1+D_1^2\otimes1+1\otimes D_2^2)^{-(p_1+p_2)/2}\leq (1+D_1^2\otimes1)^{-p_1/2}(1+1\otimes D_2^2)^{-p_2/2}.
\]
Factorisation of the trace now gives
\[
\tau\big((1+D_1^2+D_2^2)^{-(p_1+p_2)/2}\big)\leq \tau_1\big((1+D_1^2)^{-p_1/2}\big)\tau_2\big((1+D_2^2)^{-p_2/2}\big)<\infty.
\]
So the triple is $(p_1+p_2)$-$\tau$-summable. If both triples are $\theta$-$\tau_i$-summable, then
\[
\tau\big(e^{-tD^2}\big) = \tau\big(e^{-t(D_1^2\otimes1 + 1\otimes D_2^2)}\big) = \tau\big(e^{-tD_1^2}\otimes e^{-tD_2^2}\big) =  \tau_1\big(e^{-tD_1^2}\big)\,\tau_2\big(e^{-tD_2^2}\big)<\infty.
\]
Thus the triple is $\theta$-$\tau$-summable.
\end{pf}

In the case of spectral triples it has been proved \cite{Uuye} that the product of two regular spectral triples is again regular. We will now prove it in a different way (not involving pseudo-differential operators) for semifinite spectral triples. We will use some results due to Connes: Lemma \ref{domain_D_preserved} and Lemma \ref{regularity-D} below. Those results are about spectral triples and not semifinite ones, but since there is no trace or compactness of the resolvent involved, we can use them. In the following we will use the notation $\Hcal_\infty:=\bigcap_{n\in\Nat} \Dom(D^n)$.

\begin{Lemma}\label{domain_D_preserved}
Suppose $D$ is a self-adjoint operator and the operators $a,\delta(a):\Hcal_\infty\ra\Hcal$ are bounded. Then $a$ preserves $\Dom(|D|) =\Dom(D)$ and on $\Dom(D)$ the bounded extension of $\delta(a)=[|D|,a]$ and the commutator $|D|T-T|D|$ coincide.
\end{Lemma}
\begin{pf}
See \cite[Lemma 13.1]{Connes4}.
\end{pf}

\begin{Lemma}\label{regularity-D}
Suppose $(\Acal,\Hcal,D)$ is a spectral triple and $a:\Hcal_\infty\ra\Hcal$ is a bounded linear operator. Denote $\delta_1(a):= [D^2,a](1+D^2)^{-1/2}$. Then the following holds:
\begin{enumerate}[label=(\roman*)]
\item If $\delta_1(a)$ and $\delta_1^2(a)$ are bounded, then $\delta(a)$ is bounded;
\item The operators $\delta_1^n(a)$ are bounded for all $n\geq1$ if and only if $\delta^n(a)$ are bounded for all $n\geq1$.
\end{enumerate}
\end{Lemma}
\begin{pf}
See \cite[Lemma 13.2]{Connes4}.
\end{pf}

\begin{Thm}\label{regularity_product}
Suppose that for $i=1,2$ the tupels $\Scal_i:=(\Acal_i,\Hcal_i,D_i;\Ncal_i,\tau_i)$ are regular semifinite spectral triples and $\Scal_1$ is even with grading $\gamma_1$, then the product $\Scal$ of these spectral triples is again regular.
\end{Thm}
\begin{pf}
We start by showing that $\Hcal_{\infty}={\Hcal_1}_\infty\otimes{\Hcal_2}_\infty$. Recall
\[
D^{2m}=(D_1^2\otimes1+1\otimes D_2^2)^m = \sum_{k=0}^m \binom{m}{k} D_1^{2k}\otimes D_2^{2(m-k)}.
\]
Furthermore using Lemma \ref{Dom(D_1)tensor1} one can easily show that $\Dom(D_1^n\otimes1)=\Dom(D_1^n)\otimes\Hcal_2$, a similar statement holds for $D_2$. Thus by commutativity of $D_1^k\otimes 1$ and $1\otimes D_2^j$ we have
\begin{align*}
\Dom\big(D^{2m}\big)&= \bigcap_{k=0}^m\Dom\big((D_1^{2k}\otimes1)(1\otimes D_2^{2(m-k)})\big)\\
&= \bigcap_{k=0}^m \Dom\big(D_1^{2k}\big)\otimes\Hcal_2\cap \big(D_1^{2k}\otimes1\big)^{-1}(\Hcal_1\otimes\Dom\big(D_2^{2(m-k)}\big)\\
&= \bigcap_{k=0}^m \Dom\big(D_1^{2k}\big)\otimes\Hcal_2\cap \Dom\big(D_1^{2k}\big)\otimes\Dom\big(D_2^{2(m-k)}\big)\\
&= \Dom\big(D_1^{2m}\big)\otimes\Dom\big(D_2^{2m}\big).
\end{align*}
If we now use the even case, we obtain the following result for odd powers of $D$
\begin{align*}
\Dom&\big(D^{2m+1}\big) = \Dom\big(D^{2m}(D_1\otimes1+\gamma_1\otimes D_2)\big)\\
&= \Dom\big(D_1\big)\otimes\Hcal_2\cap (D_1\otimes1)^{-1}\big(\Dom\big(D^{2m}\big)\big)\\
&\qquad\cap \Hcal_1\otimes\Dom\big(D_2\big)\cap(\gamma_1\otimes D_2)^{-1}\big(\Dom\big(D^{2m}\big)\big)\\
&= \Dom\big(D_1\big)\otimes\Dom\big(D_2\big)\cap (D_1^{-1}\otimes1)\big(\Dom\big(D_1^{2m}\big)\otimes\Dom\big(D_2^{2m}\big)\big)\\
&\qquad\cap(\gamma_1\otimes D_2^{-1})\big(\Dom\big(D_1^{2m}\big)\otimes\Dom\big(D_2^{2m}\big)\big)\\
&= \Dom\big(D_1\big)\otimes\Dom\big(D_2\big)\cap \Dom\big(D_1^{2m+1}\big)\otimes\Dom\big(D_2^{2m}\big)\cap \Dom\big(D_1^{2m}\big)\otimes\Dom\big(D_2^{2m+1}\big)\\
&= \Dom\big(D_1^{2m+1}\big)\otimes\Dom\big(D_2^{2m+1}\big).
\end{align*}
As a result we obtain  $\Hcal_{\infty}={\Hcal_1}_\infty\otimes{\Hcal_2}_\infty$.\\
Recall $\Acal=\Acal_1\otimes\Acal_2$. Since $[a_1\otimes a_2,D]=[a_1,D_1]\otimes a_2 + \gamma_1a_1\otimes[a_2,D_2]$, for regularity it is sufficient to show that $a_1\otimes a_2$ is smooth if both $a_1$ and $a_2$ are smooth (thus $a_i$ is not necessarily an element from the algebra $\Acal_i$). So suppose that $a_1$ and $a_2$ are smooth. We have to prove that for all $n\geq1$ it holds that $\delta^n(a_1\otimes a_2)(\Dom(D))\subset \Dom(D)$ and $\delta^n(a_1\otimes a_2)$ are bounded. To prove the second claim we invoke Lemma \ref{regularity-D},
\begin{align}
\delta_1&(a_1\otimes a_2) = [D^2,a_1\otimes a_2](1+D^2)^{-1/2} \notag\\
&=\big([D_1^2,a_1]\otimes a_2 + a_1\otimes[D_2^2,a_2]\big)(1+D^2)^{-1/2}\notag\\
&= \big([D_1,a_1](D_1^2+1)^{-1/2}\otimes a_2\big)\; \big((D_1^2+1)^{1/2}\otimes1\big)\,(1+D^2)^{-1/2} \notag\\
&\qquad + \big(a_1\otimes[D_2,a_2](D_2^2+1)^{-1/2}\big)\; \big(1\otimes (D_2^2+1)^{1/2}\big)\,(1+D^2)^{-1/2}\notag\\
&=(\delta_1(a_1)\otimes a_2) \big((D_1^2+1)^{1/2}\otimes1\big)\,(1+D^2)^{-1/2} + (a_1\otimes \delta_1(a_2))\big(1\otimes (D_2^2+1)^{1/2}\big)\,(1+D^2)^{-1/2}.\label{eq-regularity_product1}
\end{align}
We will show that \eqref{eq-regularity_product1} is bounded, for this we will only show that the first summand is bounded, the other one is similar. Note
\[
D_1^2\otimes1 +1\leq D_1^2\otimes1 +1\otimes D_2^2+1 =D^2+1,
\]
so
\begin{align*}
1&\leq(D_1^2\otimes1+1)^{-1/2}(D +1)(D_1^2\otimes1+1)^{-1/2};\\
1&\geq(D_1^2\otimes1+1)^{1/2}(D +1)^{-1}(D_1^2\otimes1+1)^{1/2};\\
1&\geq \big\|(D_1^2\otimes1+1)^{1/2}(D+1)^{-1/2}\big\|^2.
\end{align*}
To show that higher powers $\delta_1^n(a_1\otimes a_2)$ are bounded one can do the same as in \eqref{eq-regularity_product1}. If one expands $\delta_1^n(a_1\otimes a_2)$, one gets a sum of products of elements of the form
\[
\delta_1^k(a_1)\otimes\delta_1^l(a_2),\qquad \big((D_1^2+1)^{1/2}\otimes1\big)(1+D^2)^{-1/2},\qquad \big(1\otimes (D_2^2+ 1)^{1/2}\big)\,(1+D^2)^{-1/2}
\]
and they are all bounded. Hence by Lemma \ref{regularity-D} the operators $\delta_1^n(a_1\otimes a_2)$ are bounded for all $n\geq1$. That the operators $\delta^n(a_1\otimes a_2)$ preserve the domain of $D$ is now a direct consequence of Lemma \ref{domain_D_preserved}. We conclude that $a\in\Dom(\delta^n)$ for all $n\in\Nat$. So the triple is regular.
\end{pf}

\section{Spaces of real dimension}\label{real_spaces}
This section is based on the work of Connes and Marcolli in \cite[\textsection1.19.2]{Connes2}. In this paragraph Connes and Marcolli propose a definition of a class of semifinite spectral triples which can be considered as geometric spaces of dimension $z$ for $z\in(0,\infty)$. We will give a slightly different construction, derive properties of these triples and compare it to the triples of Connes and Marcolli. We start by giving the definition and we will derive some general properties. In the second subsection we will show that the dimension spectrum consists of the singleton $\{z\}$ such that in combination with the generalised Gaussian integral the triples can indeed be considered as geometric spaces of dimension $z$.

\subsection{A $z^+$-summable semifinite spectral triple}
We want to construct for each $z\in\Com$ a spectral triple $(\Acal,\Hcal,D_z)$ which has the following property
\begin{equation}\label{eq-Trace_Dz}
\Tr\big(e^{-\lambda D_z^2}\big) = \pi^{z/2}\lambda^{-z/2}, \qquad \textrm{for all } \lambda>0.
\end{equation}
This requirement comes from the Gaussian integral
\[
\int_{\Rea^n} e^{-\lambda p^2} \,d^np = \Big(\frac{\pi}{\lambda}\Big)^{n/2}, \textrm{ for } n\in\{1,2,3,\ldots\}.
\]
The requirement \eqref{eq-Trace_Dz} may look like an simple one, but it appears to be a great constraint (c.f. Proposition \ref{restriction_Dz}). Connes and Marcolli construct such a spectral triple, but in their construction one is are naturally led to the class of semifinite spectral triples, classical spectral triples are not sufficient. Namely in the construction an operator $Z$ is needed which has the property
\begin{equation}\label{eq-spectral_decomposition_Z}
\tau(1_E(Z))=\frac{1}{2}\int_E 1\; dx, \qquad \textrm{for all } E\subset\Rea \textrm{ Borel set},
\end{equation}
which cannot be satisfied by the ordinary trace $\Tr$, details on this construction can be found in \cite{Connes2}.

\begin{Not}\label{def_Tz}
For $z>0$ consider the following tuple
\begin{equation*}
\Tcal_z:=(\Acal_z, \Hcal_z,\Dz, \Ncal_z, \tau_z):= \Big(\Com, L^2(\Rea),\Dz; L^{\infty}(\Rea), \frac{1}{2}\int_{\Rea}\;\cdot\; dx \Big).
\end{equation*}
Here $L^{\infty}(\Rea)$ acts on $L^2(\Rea)$ by pointwise multiplication of functions. Denote
\[
f_z:\Rea\ra\Rea, \qquad x \mapsto\rho(z)\sgn(x)|x|^{1/z},
\]
where $\rho(z):= \pi^{-1/2}(\Gamma(z/2 + 1))^{1/z}$ is a normalisation constant. Now $\Dz$ is given by $\Dz h:= f_zh$ on the domain
\[
\Dom(\Dz):=\Big\{h\in L^2(\Rea)\,:\, \int_{\Rea} |f_zh|^2\,dx<\infty\Big\}.
\]
We will use $\Tcal_z$, $\Dz$ etcetera in the rest of this paper to denote this triple and elements thereof.
\end{Not}

\begin{Prop}\label{triple_Tz}
For $z>0$ the tuple $\Tcal_z$ is a semifinite spectral triple, $\Dz$ has spectrum $\sigma(\Dz)=\Rea$ and the triple satisfies
\begin{equation}\label{eq-Trace_Dz2}
\tau_z\big(e^{-\lambda \Dz^2}\big) = \pi^{z/2}\lambda^{-z/2}, \qquad \textrm{for all } \lambda>0.
\end{equation}
\end{Prop}
\begin{pf}
It is clear that $L^2(\Rea)$ is a Hilbert space. The type I von Neumann algebra $L^{\infty}(\Rea)$ acts on $L^2(\Rea)$ by left-multiplication, because $\big(\int_{\Rea} |fh|^2\,dx\big)^{1/2} \leq \|f\|_{\infty} \,\big(\int_{\Rea} |h|^2\,dx\big)^{1/2}$. Clearly the trace $\tau_z$ is faithful and semifinite. From the monotone convergence theorem of measure theory it immediately follows that $\tau_z$ is normal. Since $\Acal_z=\Com$ it is obvious that $\Acal_z\subset\Ncal_z$ and $[\Dz,a]=0$, which thus extends to a bounded operator on $\Hcal_z$. \\
We will now show that $(\Dz,\Dom(\Dz))$ is a self-adjoint operator. Since $f_z$ is real-valued $\Dz$ is a symmetric operator, thus $\Dom(\Dz)\subset\Dom(\Dz^*)$. It remains to show the converse inclusion. Suppose $g\in L^2(\Rea)$ but $g\notin\Dom(\Dz)$. Then $\int_{\Rea} |gf_z|^2\,dx=\infty$. Observe that $f_z$ is a continuous function, so $f_z$ is bounded on compact sets. In particular for each $n\in\Nat$ there exists a constant $C_n$ such that $|f|_{[-n,n]}|\leq C_n$. Put $g_n:= gf_z1_{[-n,n]}$. Then $|g_n(x)|\leq C_n |g(x)|$ and
\[
\int_{\Rea} |f_zg_n|^2\,dx = \int_{\Rea} |f_zgf_z1_{[-n,n]}|^2\,dx \leq C_n^4 \int_{\Rea} |g|^2\,dx <\infty.
\]
So $g_n\in\Dom(\Dz)$. But
\begin{align*}
\langle D_z g_n\,\|g_n\|^{-1},g\rangle &= \Big(\int_{\Rea} |gf_z1_{[-n,n]}|^2\,dx \Big)^{-1/2}\int_{\Rea} f_z g f_z1_{[-n,n]} \bar{g}\,dx\\
&= \Big(\int_{-n}^n |f_z g|^2\,dx \Big)^{1/2} \ra\infty \qquad \textrm{as } n\ra\infty.
\end{align*}
Thus $g\notin \Dom(\Dz^*)$ and $\Dz$ is self-adjoint.\\
Since $\Dz$ is self-adjoint, $\sigma(\Dz)\subset\Rea$. We will show that the converse inclusion also holds. Observe that for $z>0$ the function $f_z:\Rea\ra\Rea$ is continuous, bijective and strictly increasing. Since $(\Dz-\lambda) h = (f_z-\lambda)h$, the only possible candidate for $(\Dz-\lambda)^{-1}$ is given by $h\mapsto (f_z-\lambda)^{-1}h$. But if $\lambda\in\Rea$ the function $f_z-\lambda$ has a zero. By continuity of $f_z$ the function $(f_z-\lambda)^{-1}$ is therefore not essentially bounded. Hence $h\mapsto (f_z-\lambda)^{-1}h$ is an unbounded map and thus $\Dz-\lambda$ is not invertible, which implies that $\sigma(\Dz)=\Rea$. \\
Since $\Dz$ is self-adjoint we can give a spectral decomposition of $\Dz$. Observe that the inverse of $f_z$ is given by
\[
f_z^{-1}(x) = \sgn(x)\Big(\frac{|x|}{\rho(z)}\Big)^{z}.
\]
Define
\[
E:\Bcal(\Rea)\ra B(L^2(\Rea)), \qquad E(A)h:= 1_{\{f_z^{-1}(A)\}}h.
\]
It is clear that $E$ is a spectral measure. For an interval $I=[f_z(a), f_z(b)]$ we have
\[
\int_{f_z(a)}^{f_z(b)} f_z(a)\,dE = f_z(a)E([f_z(a),f_z(b)]) = f_z(a)1_{[a,b]}.
\]
If we approximate the identity map $id:\Rea\ra\Rea$ and use the above identity, we see that $\Dz = \int x \,dE$, thus $E$ is the spectral measure for $\Dz$. For each $A\in\Bcal(\Rea)$, the Borel subsets of $\Rea$, it holds that $E(A)\in L^{\infty}(\Rea)=\Ncal_z$, thus $\Dz$ is affiliated with $\Ncal_z$. Also
\[
\tau(E([-\lambda,\lambda])) = \frac{1}{2}\int_{\Rea}1_{f_z^{-1}([-\lambda,\lambda])}\, dx = \frac{1}{2}\int_{f_z^{-1}(-\lambda)}^{f_z^{-1}(\lambda)} 1\,dx <\infty,
\]
hence by Theorem \ref{thm-tau-discrete}, $\Dz$ is compact relative to $\Ncal_z$. Thus $\Tcal_z$ is a semifinite spectral triple.\\
We will now show that $\Tcal_z$ satisfies the property \eqref{eq-Trace_Dz2}. For $\lambda>0$ we have
\[
\tau(e^{-\lambda \Dz^2})= \frac{1}{2}\;\int_{\Rea} e^{-\lambda\rho(z)^2|x|^{2/z}}\, dx = \int_0^\infty e^{-\lambda\rho(z)^2x^{2/z}}\, dx.
\]
Use the substitution $u=\lambda\rho(z)^2 |x|^{2/z}$, then
\[
x=\rho(z)^{-z} \Big(\frac{u}{\lambda}\Big)^{z/2}, \qquad \frac{dx}{du} = \rho(z)^{-z} \frac{z}{2}\,\lambda^{-z/2}u^{z/2-1}.
\]
So we obtain
\begin{align*}
\tau(e^{-\lambda \Dz^2}) &=\rho(z)^{-z}\lambda^{-z/2}\,\frac{z}{2}\;\int_{0}^{\infty}e^{-u}u^{z/2-1}\,du\\
&=\rho(z)^{-z}\lambda^{-z/2}\Gamma(z/2+1)\\
&=\pi^{z/2}\lambda^{-z/2}.
\end{align*}
In the last line we inserted the definition of $\rho(z)$. Hence \eqref{eq-Trace_Dz2} holds.
\end{pf}

Of course the triple constructed by Connes and Marcolli satisfies the requirement \eqref{eq-Trace_Dz}. Since the Dirac operator $\tilde{D}_z$ is given as a function of $Z$ and the spectral measure $E$ of $Z$ is known (c.f. \eqref{eq-spectral_decomposition_Z}), one can calculate $\Tr_{\Ncal}(e^{-\lambda \tilde{D}_z^2})$ via Theorem \ref{integration_trace}. Namely
\[
\Tr_{\Ncal}(e^{-\lambda \tilde{D}_z^2}) = \int_{\Rea} e^{-\lambda\rho(z)^2|x|^{2/z}}\, d\mu_{\tau,E}(x) = \frac{1}{2}\;\int_{\Rea} e^{-\lambda\rho(z)^2|x|^{2/z}}\, dx.
\]
Now the calculation is the same as in Proposition \ref{triple_Tz}. This reduction of Connes and Marcolli's spectral triple to the triple $\Tcal_z$ can be applied every time. So all results we obtain also hold for their triple.

\begin{Rem}
For $z\notin\Rea$ the operator $\Dz$ is not self-adjoint. This is easy to see because for $z\notin\Rea$ the Lebesgue measure of $\{x\,:\,f_z(x)\notin\Rea\}$ is strictly positive. Hence it does not hold that $f_z=\overline{f_z}$ almost everywhere. But then $\Dz^{*}\neq\Dz$. So for $z\notin\Rea$ the tuple $\Tcal_z$ is not a spectral triple. In fact, as the next result shows, we cannot expect that any self-adjoint operator satisfies \eqref{eq-Trace_Dz} for $z\in\Com\setminus(0,\infty)$.
\end{Rem}

\begin{Prop}\label{restriction_Dz}
Suppose $\Ncal\subset B(\Hcal)$ is a semifinite von Neumann algebra, with a faithful semifinite normal trace $\tau$. Suppose $z\notin(0,\infty)$. If $D$ is a self-adjoint (unbounded) operator on $\Hcal$ affiliated with $\Ncal$, then there exists a scalar $\lambda>0$ such that $\tau\big(e^{-\lambda D^2}\big) \neq \pi^{z/2}\lambda^{-z/2}$.
\end{Prop}
\begin{pf}
Let $\Ncal$, $\tau$, $z$ and $D$ be as in the assumptions of the proposition. Let $\lambda>0$ and consider the function
\[
f:\Com\ra\Com,\; f(w):=e^{-\lambda w^2}.
\]
Then $f(\Rea)\subset [0,1]$, in particular $f$ is real valued and bounded. Thus $e^{-\lambda D^2}$ is bounded and self-adjoint. Since $D$ is affiliated with $\Ncal$ it holds that
\[
e^{-\lambda D^2} \in \{D\}''\subset\Ncal.
\]
Therefore $\tau(e^{-\lambda D^2})$ is well-defined and it holds that $\tau\big(e^{-\lambda D^2}\big)\in [0,\infty]$ for all $\lambda>0$. Now suppose $z\notin\Rea$, then it is impossible that
\[
\Big(\frac{\pi}{\lambda}\Big)^{z/2}\in[0,\infty] \qquad\textrm{for all } \lambda>0.
\]
This proves the statement for $z\in\Com\setminus\Rea$. If we have $z\in (-\infty,0]$, let $t>s>0$. Then clearly
\begin{equation}\label{eq-restriction_Dz1}
t^{-z/2}\geq s^{-z/2}.
\end{equation}
But also $-tD^2<-sD^2$ and thus $e^{-tD^2}<e^{-sD^2}$. If $\tau\big(e^{-\lambda D^2}\big) = \pi^{z/2}\lambda^{-z/2}$ for all $\lambda$ this would give
\[
\pi^{z/2}t^{-z/2}=\tau\big(e^{-tD^2}\big)<\tau\big(e^{-sD^2}\big)=\pi^{z/2}s^{-z/2},
\]
which is a contradiction with \eqref{eq-restriction_Dz1}.
\end{pf}

A last observation about this spectral triple.

\begin{Rem}\label{D_z_grading}
The map $\Rea\ra\Rea$, $x\mapsto -x$ induces an operator on the triple $\Tcal_z$ by
\[
\gamma_z:L^2(\Rea)\ra L^2(\Rea), \qquad \gamma_z(f)(x):=f(-x).
\]
Since $\Acal_z=\Com$, the operator $\gamma_z$ clearly commutes with $\Acal$. Also it holds that
\[
\gamma_z(D_zf)(x)= D_z(-x)f(-x)=\rho(z)\sgn(-x)|-x|^{1/z}f(-x) = -D_z(\gamma_z f)(x).
\]
Clearly $\gamma_z$ is bounded and that $f\in\Dom(D_z)$ if and only if $\gamma_z f\in\Dom(D_z)$. But since $\gamma_z$ is not given by a function, it is no element of $\Ncal_z=L^\infty(\Rea)$. So $\gamma_z$ is not a grading as in Definition \ref{semifinite_def}, but it is very similar.
\end{Rem}

\subsection{Dimension spectrum}
In this subsection we will establish some facts about the dimension spectrum of the triple $\Tcal_z$. We will calculate in various ways the dimension spectrum because the function $f_z$ has a zero, which causes some problems.

\begin{Rem}\label{cutoff}
Connes and Marcolli use the set of poles of the meromorphic extension of $s\mapsto \tau(b|D|^{-s})$ for $b\in\Bcal$ as definition of the dimension spectrum. Since for this triple $\Bcal=\Com$ it is enough to compute $\tau_z\big(|\Dz|^{-s}\big)= \tau_z\big((\Dz^2)^{-s/2}\big)$. The operator $(\Dz^2)^{-s/2}$ is given by multiplication with the function $x\mapsto\rho(z)^{-s}|x|^{-s/z}$. So
\begin{align*}
\tau_z((\Dz^2)^{-s/2}) &= \frac{1}{2}\;\int_{\Rea}\rho(z)^{-s}|y|^{-s/z}\,dy\\
&= \rho(z)^{-s}\int_{0}^{\infty} y^{-s/z}\,dy\\
&= \rho(z)^{-s}\Big(\lim_{y\ra\infty}\big(y^{(-s+z)/z}\frac{z}{-s+z}\big) - \lim_{y\ra 0}\big(y^{(-s+z)/z}\frac{z}{-s+z}\big)\Big).
\end{align*}
But this yields $\infty$ for all values of $s$. To solve this problem an infrared cutoff is imposed, that is the integral is computed on the subset $(-\infty,1]\cup[1,\infty)$ instead of on $\Rea$. Then for $\re(\frac{s}{z})>1$ we have
\begin{equation}\label{eq-cutoff3}
\rho(z)^{-s}\int_{1}^{\infty} y^{-s/z}\,dy = \rho(z)^{-s} \Big(\lim_{y\ra\infty}\big(y^{(-s+z)/z}\frac{z}{-s+z}\big) - 1^{(-s+z)/z}\frac{z}{-s+z}\Big) = \rho(z)^{-s}\frac{z}{s-z}.
\end{equation}
This function has a meromorphic continuation to $\Com$ with a simple pole for $s=z$. The residue of \eqref{eq-cutoff3} at $s=z$ is given by
\begin{equation}\label{eq-cutoff4}
\res_{s=z}\rho(z)^{-s}\frac{z}{s-z} = z\big(\pi^{-1/2}(\Gamma(z/2+1))^{1/z}\big)^{-z} = 2\,\frac{\pi^{z/2}}{\Gamma(\frac{z}{2})}.
\end{equation}
Connes and Marcolli also gave an alternative to such a cutoff, namely to smoothly change the function $f_z$ such that the new function does not attain $0$. We will work out their suggestion. The function $g:[0,\infty)\ra[0,1]$ given by
\[
g(x):=\begin{cases}
0 &\textrm{ if } x\geq\frac{1}{2}\\
e^{-1/(x-\frac{1}{2})}  &\textrm{ if } x<\frac{1}{2}
\end{cases},
\]
is smooth and strictly decreasing. The interval $[0,\frac{1}{2}]$ is compact, $g'$ is continuous, so $g'$ is bounded on $[0,\frac{1}{2}]$. Let $c>0$ be such that $cg'(x)\leq-\frac{1}{2}$ for all $x\in[0,\frac{1}{2}]$. Define $f:[0,\infty)\ra[0,\infty)$, $f(x):=x+cg(x)$, then
\begin{itemize}
\item $f(x)=x$ if $x\geq\frac{1}{2}$;
\item $f'(x)\geq\frac{1}{2}$ for all $x\in[0,\infty)$, thus $f$ is strictly increasing;
\item $f(0)=0+cg(0)>0$.
\end{itemize}
Define an operator $E_{z}$ on $\Hcal$ by $E_{z}h(x):= \rho(z)\sgn(x)f(|x|)^{1/z}h(x)$, it is a smooth modification of $\Dz$ near $x=0$.
\end{Rem}

This operator $E_z$ is very closely related to $\Dz$, therefore we expect that the meromorphic continuation of $s\mapsto \tau_z(|E_z|^{-s})$ has the same poles as the meromorphic continuation of \eqref{eq-cutoff3}, the infra-red cutoff of $\tau_z(|\Dz|^{-s})$. This is indeed the case.

\begin{Prop}\label{dim_spec_Ez}
The meromorphic continuation of $s\mapsto\tau_z(|E_z|^{-s})$ is holomorphic on $\Com\setminus\{z\}$ and has a simple pole at $s=z$ with residue
\[
\res_{s=z}\tau_z(|E_z|^{-s})= 2\,\frac{\pi^{z/2}}{\Gamma(\frac{z}{2})}.
\]
\end{Prop}
\begin{pf}
From the construction of $E_z$ it follows that
\begin{align*}
\tau\big(|E_z|^{-s}\big) &= \int_0^\infty \rho(z)^{-s} f(x)^{-s/z} \,dx\\
&= \rho(z)^{-s} \int_0^{\frac{1}{2}} f(x)^{-s/z}\, dx + \rho(z)^{-s} \int_{\frac{1}{2}}^\infty x^{-s/z}\, dx\\
&= \rho(z)^{-s} \int_0^{\frac{1}{2}} f(x)^{-s/z}\, dx + \rho(z)^{-s} \, \frac{z}{z-s}\, \Big(\frac{1}{2}\Big)^{-s/z+1}.
\end{align*}
Observe that $f$ is continuous and non-zero, so the integral $\int_0^{\frac{1}{2}} f(x)^{-s/z}\, dx$ exists for all $s$ and therefore does not create any singularities. The second term $\rho(z)^{-s} (-\frac{s}{z} + 1)^{-1} \big(\frac{1}{2}\big)^{-s/z+1}$ has precisely a simple pole at $s=z$. Therefore
\[
\res_{s=z}\tau_z(|E_z|^{-s})= \rho(z)^{z} z \Big(\frac{1}{2}\Big)^{-z/z+1} = \pi^{z/2}z \Big(\Gamma\big(\frac{z}{2}+1\big)\Big)^{-1} =
2\,\frac{\pi^{z/2}}{\Gamma(\frac{z}{2})},
\]
as desired.
\end{pf}

A modification of the Dirac operator is not necessary if one uses $(1+D^2)^{-1/2}$ in the definition of the dimension spectrum instead of $|D|^{-1}$. We will compute the dimension spectrum of $\Tcal_z$ with $(1+D^2)^{-1/2}$. For this we need the machinery of hypergeometric functions. Before we will compute the dimension spectrum we state the properties of the hypergeometric functions which we will use.

\begin{Not}
We will use the shorthand notation $F(a,b;c;z):=\, _2F_1(a,b;c;z)$ to denote the hypergeometric function. The definition and properties of this function can be found in several books, for example in \cite[Ch. 15]{Olver}. The following two identities can be found in \cite[Ch. 15]{Olver}.
\begin{align}
F(0,b;c;z)&=F(a,0;c;z)=1, \qquad \textrm{for all } a,b,c,z\in\Com; \label{eq-hypergeom01} \\
F(a,b;c;z)&=\frac{\Gamma(c)\Gamma(b-a)}{\Gamma(b)\Gamma(c-a)}(-z)^{-a}F(a,1-c+a;1-b+a;1/z) \notag\\
&\qquad+ \frac{\Gamma(c)\Gamma(a-b)}{\Gamma(a)\Gamma(c-b)}(-z)^{-b}F(b,1-c+b;1-a+b;1/z). \label{eq-hypergeom_transf1}
\end{align}
Equality \eqref{eq-hypergeom_transf1} holds if the following requirements are satisfied $|\arg(-z)|<\pi$ and $1-b+a,\;1-a+b\notin\{0,-1,-2,\ldots\}$. \\
The last property we will use is that the indefinite integral of the function $f(x)=(1+x^p)^q$  is given by
\begin{equation}\label{eq-hypergeom_integral1}
x\mapsto xF\Big(\frac{1}{p},-q;1+\frac{1}{p};-x^p\Big).
\end{equation}
This will need a proof.
\end{Not}
\begin{pf}
By the binomial theorem we have
\[
(1+x^p)^q = \sum_{k=0}^\infty \binom{q}{k}x^{pk}.
\]
Here $\binom{q}{k}$ for $q\in\Com$ is defined by $\binom{q}{k}:= \frac{(q)_k}{k!}$, with $(q)_k:=q(q-1)\cdots(q-k+1)$. Taking primitives on both sides yields
\begin{align*}
\int (1+x^p)^q &= \sum_{k=0}^\infty \binom{q}{k} \frac{1}{pk+1}\;x^{pk+1}\\
&= x\sum_{k=0}^\infty \frac{q(q-1)\cdots(q-k+1)}{k!(pk+1)}\; (x^p)^k\\
&= x\sum_{k=0}^\infty \frac{\frac{1}{p}(\frac{1}{p}+1)\cdots(\frac{1}{p}+k-1) (-q)(-q+1)\cdots(-q+k-1)}{(\frac{1}{p}+1)\cdots(\frac{1}{p}+k-1)(\frac{1}{p}+k)k!}\;(-x^p)^k\\
&= xF\Big(\frac{1}{p},-q;1+\frac{1}{p};-x^p\Big),
\end{align*}
as desired.
\end{pf}

\begin{Thm}\label{dim_spec_Dz}
Assume $z\in(0,\infty)$. Then the triple $\Tcal_z$ is $z^+$-summable, regular, the dimension spectrum is simple and consists of $\{z\}$ and
\[
\res_{s=z} \tau_z((1+\Dz^2)^{s/2}) = \frac{\pi^{z/2}}{\Gamma(z/2)}.
\]
\end{Thm}
Note that we have a factor $2$ difference between the residue at $s=z$ of $s\mapsto\tau_z((1+\Dz^2)^{-s/2})$ and \eqref{eq-cutoff4}.

\smallskip
\begin{pf}
Since $\Acal_z=\Com$, the commutators $[\Dz,a]=[|\Dz|,a]=0$ for all $a\in\Acal_z$, thus it is obvious that the triple is regular. For the dimension spectrum we have to compute the poles of $s\mapsto \tau_z(b(1+\Dz^2)^{-s/2})$. We can take $b=1$, since $\Acal_z=\Com$. Thus we consider the meromorphic function
\begin{align*}
s\mapsto \tau_z((1+\Dz^2)^{-s/2})&=\frac{1}{2}\,\int_{\Rea}\big(1+(\rho(z)\sgn(x)|x|^{1/z})^2\big)^{-s/2}\,dx\\
&=\int_{0}^{\infty}(1+\rho(z)^2x^{2/z})^{-s/2}\,dx\\
&=\rho(z)^{-z}\int_{0}^{\infty}(1+y^{2/z})^{-s/2}\,dy.
\end{align*}
The last equality follows from the substitution $y=\rho(z)^zx$. Note that the constant $\rho(z)^z$ does not affect the location of the poles. By Equation \eqref{eq-hypergeom_integral1} we have
\[
\int_{0}^{\infty}(1+x^{2/z})^{-s/2}\,dx= \lim_{x\ra\infty} xF\Big(\frac{z}{2},\frac{s}{2};1+\frac{z}{2};-x^{2/z}\Big).
\]
Since $2/z \in[0,\infty)$, for every $x>0$ we have $-x^{2/z}<0$. So $|\arg(--x^{2/z})|=|\arg(x^{2/z})|=0<\pi$. Therefore if $1-z/2+s/2\notin\{0,-1,-2,\ldots\}$, i.e. if $s\notin\{-2+z,-4+z,\ldots\}$ we can apply \eqref{eq-hypergeom_transf1} and \eqref{eq-hypergeom01} to obtain
\begin{align}
\lim_{x\ra\infty} &xF\Big(\frac{z}{2},\frac{s}{2};1+\frac{z}{2};-x^{2/z}\Big) \label{eq-dim_specDz4} \\
&= \lim_{x\ra\infty} \Big(x\,\frac{\Gamma(1+\frac{z}{2})\Gamma(\frac{s}{2}-\frac{z}{2})}{\Gamma(\frac{s}{2})\Gamma(1)}\, (x^{2/z})^{-z/2} F\big(\frac{z}{2},0;1-\frac{s}{2}+\frac{z}{2};(-x^{2/z})^{-1}\big) \notag\\
&\qquad +x\,\frac{\Gamma(1+\frac{z}{2})\Gamma(\frac{z}{2}-\frac{s}{2})} {\Gamma(\frac{z}{2})\Gamma(1+\frac{z}{2}-\frac{s}{2})}\,(x^{2/z})^{-s/2} F\big(\frac{s}{2},-\frac{z}{2}+\frac{s}{2};1-\frac{z}{2}+\frac{s}{2};(-x^{2/z})^{-1}\big)\Big) \notag\\
&=\frac{\Gamma(1+\frac{z}{2})\Gamma(\frac{s}{2}-\frac{z}{2})}{\Gamma(\frac{s}{2})}\notag \\
&\qquad + \lim_{x\ra\infty} \frac{\Gamma(1+\frac{z}{2})\Gamma(\frac{z}{2}-\frac{s}{2})} {\Gamma(\frac{z}{2})\Gamma(1+\frac{z}{2}-\frac{s}{2})}\,x^{-s/z+1} \, F\big(\frac{s}{2},-\frac{z}{2}+\frac{s}{2};1-\frac{z}{2}+\frac{s}{2};-x^{-2/z}\big).
\label{eq-dim_specDz1}
\end{align}
Now suppose $s\in\Com$ such that $\re(s)>z$, then $\re(-s/z+1)<0$. On the disk $\{z\in\Com\,:\, |z|<1\}$ the function $z\mapsto F(a,b;c;z)$ is holomorphic. Since $F(a,b;c;0)=1$ we have
\begin{equation}
\lim_{x\ra\infty}x^{-s/z+1} \, F\big(\frac{s}{2},-\frac{z}{2}+\frac{s}{2};1-\frac{z}{2}+\frac{s}{2};-x^{-2/z}\big) = 0\label{eq-dim_specDz2}.
\end{equation}
Inserting \eqref{eq-dim_specDz2} in \eqref{eq-dim_specDz1} gives for $\re(s)>z$
\[
\lim_{x\ra\infty} xF\big(\frac{z}{2},\frac{s}{2};1+\frac{z}{2};-x^{2/z}\big) = \frac{\Gamma(1+\frac{z}{2})\Gamma(\frac{s}{2}-\frac{z}{2})}{\Gamma(\frac{s}{2})}.
\]
By analytic continuation this holds for any $s\in\Com\setminus\ \{-2+z,-4+z,\ldots\}$. Recall that $\rho(z)=\pi^{-1/2}(\Gamma(1+\frac{z}{2}))^{1/z}$, so
\[
\tau_z\big((1+\Dz^2)^{-s/2}\big)=\frac{\Gamma(\frac{s}{2}-\frac{z}{2})}{\Gamma(\frac{s}{2})}\,\pi^{z/2}, \qquad \textrm{for } s\in\Com\setminus\ \{-2+z,-4+z,\ldots\} .
\]
Observe that the function $s\mapsto\Gamma(\frac{s}{2}-\frac{z}{2})$ has simple poles for $\frac{s}{2}-\frac{z}{2}\in\{\ldots,-2,-1,0\}$. Note that $w\mapsto \big(\Gamma(\frac{w}{2})\big)^{-1}$ is holomorphic. Thus $s\mapsto \tau_z(1+\Dz^2)^{-s/2}$ has a simple pole at $s=z$. For $m\in\Nat$ the residues of the gamma function are given by $\res_{w=-m}\Gamma(w) = \frac{(-1)^m}{m!}$. Thus it follows that
\[
\res_{s=z} \tau_z\big((1+\Dz^2)^{-s/2}\big) = \res_{s=z}\frac{\Gamma(\frac{s}{2}-\frac{z}{2})}{\Gamma(\frac{s}{2})}\,\pi^{z/2} = \frac{\pi^{z/2}}{\Gamma(\frac{z}{2})}.
\]
For every pole $s$ we have $s\leq z$. Therefore $\tau_z(1+\Dz^2)^{-s/2}<\infty$ for all $s>z$, hence the triple is $z^+$-summable.\\
It remains to show that $z$ is the only pole. Recall that the transformation \eqref{eq-dim_specDz1} was only valid for $1-z/2+s/2\notin\{0,-1,-2,\ldots\}$. We have to deal with these points in a different way. Fix $x_0\in(1,\infty)$ and $s_0\in\{\ldots, -4+z,-2+z\}$. Say $-\frac{z}{2}+\frac{s_0}{2}=-m$. Let $s$ be close to $s_0$, then
\begin{align*}
F\big(\frac{s}{2},&-\frac{z}{2}+\frac{s}{2};1-\frac{z}{2}+\frac{s}{2};-x_0^{-2/z}\big)\\
&= \sum_{j=0}^\infty \frac{(\frac{s}{2})_j (-\frac{z}{2}+\frac{s}{2})(-\frac{z}{2}+\frac{s}{2}+1)\cdots (-\frac{z}{2}+\frac{s}{2}+j-1)}{(-\frac{z}{2}+\frac{s}{2}+1)\cdots(-\frac{z}{2}+\frac{s}{2}+j-1)(-\frac{z}{2}+ \frac{s}{2}+j)j!}\, (-x_0^{-2/z})^j \\
&= \sum_{j\in\Nat\setminus\{m\}} \frac{(\frac{s}{2})_j}{j!}\,\frac{-\frac{z}{2}+\frac{s}{2}}{-\frac{z}{2}+\frac{s}{2} -j}\,(-x_0^{z/2})^j + \frac{(\frac{s}{2})_m (-\frac{z}{2}+\frac{s}{2})}{(-\frac{z}{2}+\frac{s}{2}+m)m!}\, (-x_0^{-2/z})^m.
\end{align*}
We will now consider the limit $s\ra s_0$. To prove that \eqref{eq-dim_specDz4} does not have a pole at $s=s_0$ we compute (still for $x_0$ fixed) the residue of \eqref{eq-dim_specDz4} and show that it equals $0$. Since we singled out $m$ in the infinite sum, near $s_0$ the function
\[
s\mapsto \sum_{j\in\Nat\setminus\{m\}} \frac{(\frac{s}{2})_j}{j!}\,\frac{-\frac{z}{2}+\frac{s}{2}}{-\frac{z}{2}+\frac{s}{2} -j}\,(-x_0^{-2/z})^j
\]
is holomorphic. To conclude that \eqref{eq-dim_specDz1} has a removable singularity at $s_0$ we compute the following residue
\begin{align*}
\res_{s=s_0} &\frac{\Gamma(1+\frac{z}{2})\Gamma(\frac{s}{2}-\frac{z}{2})}{\Gamma(\frac{s}{2})} + \frac{\Gamma(1+\frac{z}{2})\Gamma(\frac{z}{2}-\frac{s}{2})} {\Gamma(\frac{z}{2})\Gamma(1+\frac{z}{2}-\frac{s}{2})}\,x_0^{-s/z+1}\;\frac{(\frac{s}{2})_m (-\frac{z}{2}+\frac{s}{2})}{(-\frac{z}{2}+\frac{s}{2}+m)m!} (-x_0^{-2/z})^m\\
&= \frac{\Gamma(1+\frac{z}{2})}{\Gamma(\frac{s_0}{2})}\;\frac{(-1)^m}{m!} + \frac{\Gamma(1+\frac{z}{2})\Gamma(m)} {\Gamma(\frac{z}{2})\Gamma(m+1)}\,x_0^{-s_0/z+1}\;\frac{(\frac{s_0}{2})_m (-m)}{m!} (-1)^m(x_0^{-2/z})^m\\
&=\frac{\Gamma(1+\frac{z}{2})}{\Gamma(\frac{s_0}{2})}\;\frac{(-1)^m}{m!} - \frac{\Gamma(1+\frac{z}{2})}{\Gamma(\frac{z}{2})}\, \frac{(m-1)!m}{m!}\, \frac{1}{m!} \frac{\Gamma(\frac{s_0}{2}+m)}{\Gamma(\frac{s_0}{2})}(-1)^m\\
&= \frac{\Gamma(1+\frac{z}{2})}{\Gamma(\frac{s_0}{2})}\;\frac{(-1)^m}{m!} - \frac{\Gamma(1+\frac{z}{2})}{\Gamma(\frac{z}{2})}\;\frac{(-1)^m}{m!} \frac{\Gamma(\frac{z}{2})}{\Gamma(\frac{s_0}{2})} =0.
\end{align*}
Indeed, this residue is independent of $x_0$. So the dimension spectrum of $\Dz$ consists of $z=s$.
\end{pf}

It is interesting to compare the dimension spectrum of $\Dz$ and $E_z$, the latter operator is the operator introduced in Remark \ref{cutoff}. With the previous theorem it has become easy to compute the poles of $s\mapsto \tau_z((1+E_z^2)^{-s/2})$.

\begin{Prop}
Suppose $z\in(0,\infty)$, then the dimension spectrum of $E_z$ equals the dimension spectrum of $\Dz$.
\end{Prop}
\begin{pf}
From the definition of the trace $\tau_z$ and the operator $E_z$ it immediately follows that
\begin{align*}
\tau_z\big((1+E_z^2)^{-s/2}\big) &= \int_0^\infty \big(1+(\rho(z)\sgn(x)f(|x|)^{1/z})^2\big)^{-s/2}\,dx\\
&= \int_0^{1/2} \big(1+\rho(z)^2f(|x|)^{2/z}\big)^{-s/2}\,dx + \int_{1/2}^\infty \big(1+\rho(z)^2x^{2/z}\big)^{-s/2}\,dx.
\end{align*}
It follows from the proof of Theorem \ref{dim_spec_Dz} that it suffices to show that the functions
\begin{align*}
s\mapsto \int_0^{1/2} \big(1+\rho(z)^2x^{2/z}\big)^{-s/2}\,dx,& &
s\mapsto \int_0^{1/2} \big(1+\rho(z)^2f(|x|)^{2/z}\big)^{-s/2}\,dx
\end{align*}
are holomorphic on $\Com$. We start with the first one.
\[
\int_0^{1/2} \big(1+\rho(z)^2x^{2/z}\big)^{-s/2}\,dx = \rho(z)^{-z}\,\int_0^{1/2} (1+x^{2/z})^{-s/2}\,dx = \rho(z)^{-z}\frac{1}{2}\,F\Big(\frac{z}{2},\frac{s}{2};1+\frac{z}{2};-\big(\frac{1}{2}\big)^{2/z}\Big).
\]
Note $-(1/2)^{2/z}\in(-\infty,0)$, thus \cite[\textsection 15.2]{Olver} implies that
\[
s\mapsto \rho(z)^{-z}\frac{1}{2}F\Big(\frac{z}{2},\frac{s}{2};1+\frac{z}{2}; -\big(\frac{1}{2}\big)^{2/z}\Big)
\]
is holomorphic on $\Com$. Now the second one. The function $f$ is smooth, strictly increasing, $f(0)>0$ and $f(1/2)=1/2$. Thus there exist $\delta_1, \delta_2>0$ such that $\delta_1<1+\rho(z)^2f(x)^{2/z}<\delta_2$ for all $x\in[0,1/2]$. But from these bounds it is clear that for each $s\in\Com$ the function
\[
[0,1/2]\ra \Com, \qquad x\mapsto \big(1+\rho(z)^2f(|x|)^{2/z}\big)^{-s/2}
\]
is bounded. Thus
\[
s\mapsto \int_0^{1/2} \big(1+\rho(z)^2f(|x|)^{2/z}\big)^{-s/2}\,dx
\]
does not have any poles in $\Com$.
\end{pf}

In Theorem \ref{regularity_product} it was proved that the product of two regular semifinite spectral triples is again regular. Since $\Acal_z=\Com$ and thus $\Tcal_z$ is regular, the following corollary is immediate.

\begin{Cor}
Suppose $\Scal$ is an even regular semifinite spectral triple. Then the product triple $\Scal\times\Tcal_z$ is also regular.
\end{Cor}

Since the product of a spectral triple with $\Tcal_z$ is regular, one can compute its dimension spectrum. We expect that the whole spectrum shifts over the vector $z$, but we cannot prove this fact in its full generality. The difficulty lies in computing the algebra $\Bcal$ of the product triple, because the operator $|D_1\otimes1+\gamma_1\otimes\Dz|$ is difficult to work with. However we can prove the result for the pole with the largest real part. We will use Fubini's theorem for traces c.f. Proposition \ref{interchange_trace_int}.

\begin{Prop}\label{dimspec_product}
Suppose $\Scal:=(\Acal,\Hcal,D; \Ncal,\tau,\gamma)$ is an even finitely summable regular semifinite spectral triple with $1\in\Acal$. Denote $\Scal_z:= (\Acal,\tilde{\Hcal},D_z;\tilde{\Ncal},\tau') := \Scal\times\Tcal_z$. Suppose $w\in Sd(\Scal)$, the dimension dimension spectrum of $\Scal$, such that $\re(w)>0$ and for all $w'\in Sd(\Scal)$ it holds $\re(w')\leq\re(w)$. Then if $0<z<\re(w)$, the function $s\mapsto \tau'((D_z^2+1)^{-s/2})$ has a pole at $s=w+z$ and all other singularities $w''$ of the zeta functions $\{\zeta_b\,:\,b\in\Bcal(\Scal_z)\}$ satisfy $\re(w'')\leq\re(w)+z$.
\end{Prop}
\begin{pf}
The main idea of the proof of this proposition is to write the operator $(D_z+1)^{-s/2}$ as an elementary tensor and then use the factorisation of the trace $\tau'$. This can be done by writing this operator as an integral of exponential functions and we will use Lemma \ref{interchange_trace_int} to interchange the integral and the trace. We start with the identity
\[
\int_0^\infty e^{-tx}t^{a-1}\,dt= x^{-a}\Gamma(a), \qquad \re(a)>0.
\]
In the following calculation we will interchange two times an integral with a trace. We will justify those manipulations later. For $s\in\Com$ such that $\re(s),\,\re(s-z)>0$ it holds that
\begin{align}
\zeta_1(s) &= \tau'((D_z^2 +1)^{-s/2})= \tau'\Big(\frac{1}{\Gamma(s/2)}\int_0^\infty e^{-t(1+D^2\otimes1+1\otimes \Dz^2)} t^{s/2-1}\,dt\Big)\notag\\
&= \frac{1}{\Gamma(s/2)}\int_0^\infty \tau'\big(e^{-t(1+D^2)}\otimes e^{-t\Dz^2}\big) t^{s/2-1}\,dt\label{eq-dimspec_product1}\\
&= \frac{1}{\Gamma(s/2)}\int_0^\infty \tau\big(e^{-t(1+D^2)}\big)\, \tau_z\big(e^{-t\Dz^2}\big) t^{s/2-1}\,dt\notag\\
&= \frac{1}{\Gamma(s/2)}\int_0^\infty \tau\big(e^{-t(1+D^2)}\big)\, \pi^{z/2}t^{-z/2}t^{s/2-1}\,dt\notag\\
&= \pi^{z/2}\,\frac{1}{\Gamma(s/2)}\; \tau\Big(\int_0^\infty e^{-t(1+D^2)}\, t^{(s-z)/2-1}\,dt\Big)\label{eq-dimspec_product2}\\
&= \pi^{z/2}\,\frac{\Gamma((s-z)/2)}{\Gamma(s/2)}\;\tau\big((1+D^2)^{-(s-z)/2}\big).\notag
\end{align}
By assumption the function $s\mapsto \tau((1+D^2)^{-s/2})$ has a pole at $w$, thus $s\mapsto\tau'(D_z^2 +1)^{-s/2})$ has a pole at $s=w+z$. Since $z<\re(w)$, this is the largest pole of $\zeta_1$, because $\Gamma$ only has poles at the non-positive integers. \\
Suppose $b\in\Bcal(\Scal_z)$. The following estimate shows that one does not obtain any poles in the half-plane $\{s\in\Com\,:\,\re(s)>\re(w)=z\}$.
\[
|\zeta_b(s)|=|\tau'\big(b(D_z^2+1)^{-s/2}\big)|\leq\|b\| \tau'(|(D_z^2+1)^{-s/2}|),
\]
which converges for $s$ with $\re(s)>\re(w)+z$.\\
It remains to show why one can interchange the integral and trace in \eqref{eq-dimspec_product1} and \eqref{eq-dimspec_product2} we use Proposition \ref{interchange_trace_int}. Suppose $E$ is an unbounded self-adjoint operator on $\Kcal$. By the uniqueness of the analytic continuation it is sufficient to prove that the switch is allowed for $s\in\Rea$ with $s>2$. Fix such an $s$. Consider
\[
f:[0,\infty)\ra B(\Kcal), \qquad t\mapsto \frac{1}{\Gamma(s)}\,e^{-t(1+E^2)}t^{s/2-1}.
\]
We check the conditions of Proposition \ref{interchange_trace_int}. We know that if $a,b$ are positive, then the map $[0,\infty)\ra\Rea$, $t\mapsto e^{-ta}t^b$ is uniformly bounded. Using the functional calculus $f(E)$ is norm-bounded.\\
To prove the second requirement note that $f(\cdot)h \leftrightsquigarrow f(\cdot)^*h$ corresponds to $s\leftrightsquigarrow \overline{s}$. Thus it is sufficient to show that $f(\cdot)h$ is measurable. We will prove continuity of $f(\cdot)h$.
\begin{align*}
&\|f(t_0+t)h-f(t_0)h\| \leq \frac{1}{\Gamma(s)}\,\big\|e^{(-t_0+t)(1+E^2)}(t_0+t)^{s/2-1} - e^{-t_0(1+E^2)}(t_0+t)^{s/2-1}\big\|\,\|h\|\\
&\leq \frac{1}{\Gamma(s)}\,\Big(\big\|e^{-t_0(1+E^2)}\big\|\, \big\|e^{-t(1+E^2)}-1\big\|\,|t_0+t|^{s/2-1} + \big\|e^{-t_0(1+E^2)}\big\|\, \big|(t_0+t)^{s/2-1}-t_0^{s/2-1}\big|\Big)\|h\|
\end{align*}
which tends to $0$ as $t\ra0$. For the last requirement note that
\begin{equation}\label{eq-dimspec_product3}
\rho(|f(t)|) = \frac{1}{\Gamma(s)}\rho\big(e^{-t(1+E^2)}\big)\big|t^{s/2-1}\big|.
\end{equation}
Then because $s>2$, it follows that $\lim_{t\ra 0} \rho(|f(t)|) =0$ and also $\lim_{t\ra\infty} \rho(|f(t)|) =0$. Therefore $\rho(|f(t)|)$ is uniformly bounded in $t$. If we now let $E=D_z$ and $\rho=\tau'$ then it is obvious that as a function of $t$, \eqref{eq-dimspec_product3} is uniformly bounded on $[0,\infty)$, in particular on the interval $[0,n]$. Thus using the fact that $[D^2\otimes1,1\otimes\Dz^2]=0$ and Proposition \ref{interchange_trace_int} we obtain that for all $n>0$:
\begin{equation*}
\tau'\Big(\frac{1}{\Gamma(s)}\int_0^n e^{-t(1+D^2\otimes1+1\otimes \Dz^2)} t^{s/2-1}\,dt\Big) = \frac{1}{\Gamma(s)}\int_0^n \tau'\big(e^{-t(1+D^2)}\otimes e^{-t\Dz^2}\big) t^{s/2-1}\,dt.
\end{equation*}
Taking the limit $n\ra\infty$ gives the desired Equality \eqref{eq-dimspec_product1}. For \eqref{eq-dimspec_product2}, we can do the same trick, but we have to replace $E$ by $D$ and $\rho$ by $\tau$.
\end{pf}

\section{Application to Quantum Field Theory}\label{QFT}
In section \ref{real_spaces} we described a set of spectral triples which can be considered as being $z$-dimensional. In this section we will apply these spectral triples to dimensional regularisation and zeta function regularisation in quantum field theory. We will not develop a general theory, but we will describe an example which will give a good idea how one can apply these triples in more general computations.

\bigskip
The basic idea of regularisation and renormalisation is the following. Suppose we are given a divergent expression $A$. For {\it regularisation} one modifies $A$ by inserting in some way an extra parameter $s$ to a obtain a different expression $A(s)$. We want that $A(s)$ satisfies the following two properties: $A(s)$ must be well-defined on $U\setminus\{0\}$ (where $U$ is some neighbourhood of $0$) and $A(0)=A$. The parameter $s$ is called the {\it regulator}. Since $A(0)=A$, the expression $A(s)$ will then have a pole for $s=0$. {\it Renormalisation} of $A$ is to subtract this pole at $s=0$. So
\[
A_{\textrm{ren}}:=\big(A(s)-\frac{1}{s}\,\res_{w=0} A(w)\big)|_{s=0}
\]
is then a finite quantity. In quantum field theory, this quantity $A_{\textrm{ren}}$ is what one is interested in.

\subsection{Dimensional regularisation}\label{Dim_reg}
In quantum field theory, the integrals considered are typically of the form
\begin{equation}\label{eq-dimreg1}
\int_{\Rea^4} \frac{1}{(k^2+m^2)^2} d^4k.
\end{equation}
This specific integral corresponds to the following Feynman diagram:

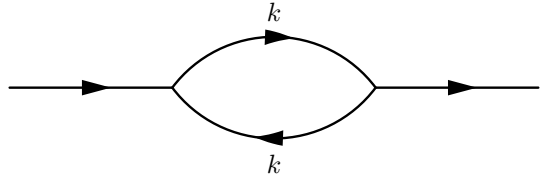
\begin{center}
\begin{fmffile}{graph1}
\begin{fmfgraph*}(200,60)
\fmfleft{i1}
\fmfright{o1}
\fmf{fermion}{i1,v1}
\fmf{fermion,left=0.5,tension=0.4,label=$k$}{v1,v2}
\fmf{fermion,left=0.5,tension=0.4,label=$k$}{v2,v1}
\fmf{fermion}{v2,o1}
\end{fmfgraph*}
\end{fmffile}
\end{center}

It represents a particle with mass $m$ which propagates and self-interacts. Since these integrals are divergent, one needs to regularize these integrals before one is able make sense of it. To regularize such integrals 't Hooft and Veltman \cite{Hooft} proposed dimensional regularisation. The idea is that if an integral $A$ is divergent in $d$ dimensions one tries to compute $A(\eps)$, the same integral but then in $d-\eps$ dimensions. The method of 't Hooft and Veltman is based on the following formula:
\begin{equation}\label{eq-dimreg2}
\int_{\Rea^D} e^{-\lambda p^2} \,d^Dp = \Big(\frac{\pi}{\lambda}\Big)^{D/2}.
\end{equation}
A priori this equality is valid for $D\in\{1,2,\ldots\}$. For non-integer values of $D$ this cannot be proved, because their is no such thing as a (Lebesgue) integral in $D$-dimensions. But instead of proving it, the right hand side is used as a definition for the left hand side if $D\notin \{1,2,\ldots\}$. In this way one obtains a method to integrate in $D$ dimensions. We will work out an example.

\begin{Exam}\label{Exam_dimreg1}
We will compute \eqref{eq-dimreg1} in dimension $4-w$ for $\re(w)>0$. To start, note that
\[
\int_0^\infty e^{-t(p^2+m^2)}\,dt = \frac{1}{p^2+m^2}.
\]
So
\begin{align}
\frac{1}{(p^2+m^2)^2} &= \int_0^\infty \int_0^\infty e^{-s(p^2+m^2)} e^{-t(p^2+m^2)}\,ds\,dt \notag\\
&= \int_0^1\int_0^\infty e^{-\lambda(p^2+m^2)}\lambda\,d\lambda\,dx \notag\\
&= \int_0^\infty e^{-\lambda(p^2+m^2)}\lambda\,d\lambda, \label{eq-Exam_dimreg1}
\end{align}
where we used the substitution $s=(1-x)\lambda$ and $t=x\lambda$. We therefore obtain
\[
\int \frac{1}{(p^2+m^2)^2}\,d^{4-w}p = \int \int_0^\infty e^{-\lambda(p^2+m^2)}\lambda\,d\lambda\, d^{4-w}p.
\]
Although we do not have a theorem of Fubini at our disposal, we just change the order of ``integration''. We do this, because this is a way to give a meaning to the integral in $4-w$-dimensions. After this we insert the essential identity \eqref{eq-dimreg2} to obtain
\begin{align*}
\int \frac{1}{(p^2+m^2)^2}\,d^{4-w}p &= \int_0^\infty \int e^{-\lambda(p^2+m^2)}\lambda\, d^{4-w}p\, d\lambda\\
&= \int_0^\infty \Big(\int  e^{-\lambda p^2}\, d^{4-w}p\Big)\lambda e^{-\lambda m^2}\,d\lambda\\
&= \int_0^\infty \Big(\frac{\pi}{\lambda}\Big)^{(4-w)/2}\,\lambda e^{-\lambda m^2}\,d\lambda\\
&= \pi^{(4-w)/2} \int_0^\infty \Big(\frac{\mu}{m^2}\Big)^{-1+w/2} e^{-\mu}\,\frac{1}{m^2}\,d\mu\\
&= \pi^{(4-w)/2} m^{-w} \int_0^\infty e^{-\mu} \mu^{-1+w/2}\,d\mu\\
&= \pi^{(4-w)/2} m^{-w} \Gamma\Big(\frac{w}{2}\Big).
\end{align*}
\end{Exam}

The general theory can for example be found in \cite[Chapter 7]{Folland}. What one usually does is introduce new variables (in our example $x$ and $\lambda$), rewrite the integrand as an exponential function and then use identity \eqref{eq-dimreg2}.\\
We however do not need to use \eqref{eq-dimreg2} as a definition, but using our previous developed machinery we can explicitly compute
\[
\tau_z\big(e^{-\lambda\Dz}\big) = \Big(\frac{\pi}{\lambda}\Big)^{z/2}
\]
and use this as a definition of an integral in $z$ dimensions instead. So if we want to calculate an integral in $z$ dimensions, we have to replace the variable over which we integrate by the operator $\Dz$ and the integral by the trace $\tau_z$. Via this method we have a genuine calculation and not just a formal manipulation. We will illustrate this with an example, we compute again \eqref{eq-dimreg1} but now with $\Dz$ and $\tau_z$.

\begin{Exam}
We have
\[
\int \frac{1}{(k^2+m^2)^2}\,d^zk := \tau_z\Big(\frac{1}{(\Dz^2+m^2)^2}\Big).
\]
As before we use identity \eqref{eq-Exam_dimreg1}, interchange integral and trace and finally we use \eqref{eq-Trace_Dz2} to obtain
\begin{align*}
\tau_z\Big(\frac{1}{(\Dz^2+m^2)^2}\Big) &= \tau_z\Big(\int_0^\infty e^{-\lambda\Dz^2}e^{-\lambda m^2}\lambda\,d\lambda\Big)\\
&= \int_0^\infty \tau_z\big(e^{-\lambda\Dz^2}\big) e^{-\lambda m^2}\lambda\,d\lambda\\
&= \int_0^\infty \Big(\frac{\pi}{\lambda}\Big)^{z/2} e^{-\lambda m^2}\lambda\,d\lambda.
\end{align*}
Note that it is valid to interchange trace and integral, because by the definition of $\tau_z$ and $\Dz$ we have
\[
\tau_z\Big(\int_0^\infty e^{-\lambda\Dz^2}e^{-\lambda m^2}\lambda\,d\lambda\Big) = \int_0^\infty\int_0^\infty e^{-\lambda \rho(z)^2x^{2/z}} e^{-\lambda m^2}\lambda\,d\lambda\,dx.
\]
Since $z>0$ it holds that $e^{-\lambda \rho(z)^2x^{2/z}} e^{-\lambda m^2}\lambda \geq 0$ for all $x,\lambda\in[0,\infty)$. Thus by Fubini interchanging the integrals is allowed and therefore we can interchange $\tau$ and the integral.\\
To finish the calculation, we can copy the end of Example \ref{Exam_dimreg1}. So
\[
\tau_z\Big(\frac{1}{(\Dz^2+m^2)^2}\Big) = \pi^{z/2}m^{4-z}\Gamma(2-z/2),
\]
which has precisely a simple pole at $z=4$. Note that this expression is well defined for all $z>0$ and has a meromorphic extension to $\Com$.
\end{Exam}

\subsection{Zeta function regularisation}
In this subsection we will show that it is also possible to use semifinite spectral triples for zeta function regularisation \cite{Hawking}. In zeta function regularisation on typically has to deal with divergent integrals of the form
\[
\int_0^\infty \frac{1}{t}\, e^{-t}\,dt.
\]
For zeta function regularisation one inserts an extra power $t^s$ to make the integral convergent and then computes the behaviour as $s\ra 0$. In the above case one gets
\[
\int_0^\infty t^{-1+s}e^{-t}\,dt = \Gamma(s),
\]
which has a simple pole at $s=0$ with residue $1$. Note that this result is very similar to dimensional regularisation: the introduced regulator $s$ appears in a gamma function which has a pole at $0$ and this pole describes exactly the divergence of the original integral.

\bigskip
Assume that we have a space given by a spectral triple $\Scal:=(\Acal,\Hcal,D;\gamma)$. In this subsection we will consider the tensor product of the triple $\Scal$ with the semifinite triple $\Tcal_z$. Denote as before
\[
\Scal_z:=(\Acal,\tilde{\Hcal},D_z;\tilde{\Ncal},\tau'):=\Scal\times\Tcal_z
\]
for the product of the spectral triple $\Scal$ with the semifinite spectral triple $\Tcal_z$ as described in Theorem \ref{semifinite_product}. The use of $\Acal$ instead of $\tilde\Acal$ is no typo, because $\Acal\otimes\Com\cong\Acal$.

\begin{Def}
Define the {\it gauge potentials} of a spectral triple $(\Acal,\Hcal,D)$ by
\[
\Omega_D^1(\Acal):=\Big\{ \sum_j a_j[D,b_j]\,:\, a_j,b_j\in\Acal\Big\}.
\]
\end{Def}

In quantum field theory one is interested in computing $\det((D+A)D^{-1})$, where $D$ is the Dirac operator of a spectral triple $(\Acal,\Hcal,D)$ and $A$ a gauge potential. However $\det$ is not defined for (un)bounded operators on infinite dimensional Hilbert spaces. Recall from linear algebra the identity $\det(\exp(T))=\exp(\Tr(T))$ for any square matrix $T$. Thus using functional calculus we have for positive square matrices $\log(\det(T))=\Tr(\log(T))$. A formal manipulation gives
\[
\log\big(\det((D+A)D^{-1})\big) = \Tr\big(\log((D+A)D^{-1})\big) = \Tr\big(\log(1+AD^{-1})\big).
\]
Therefore we are interested in computing $\Tr\big(\log(1+AD^{-1})\big)$ instead of $\det((D+A)D^{-1})$. However, in general this quantity is not finite, as the operators considered are in general not trace class. Therefore it is necessary to regularize and renormalise this quantity. We will describe a method which makes use of the semifinite spectral triple $\Tcal_z$ and the product of semifinite spectral triples. Motivated by zeta-function regularization we insert an extra factor $|D|^{-s}$ and examine in what way the result diverges as a function of $s$. So we consider the regularised quantity
\[
A(s):= \Tr\big(\log(1+AD^{-1})|D|^{-s}\big).
\]
To renormalise this, one needs to compute the residue at $s=0$. For this we use the noncommutative integral
\[
\ncint P:= \res_{s=0} \Tr(P|D|^{-s}).
\]
This functional defines a trace on the algebra generated by $\Acal, [D,\Acal]$ and $|D|^{s}$ with $s\in\Com$, if the triple has a simple discrete dimension spectrum \cite{Connes3}. The following theorem by Connes and Chamseddine about spectral actions will be useful later on.

\begin{Thm}[Connes \& Chamseddine \cite{Connes5}]\label{variation_zeta}
Suppose $(\Acal,\Hcal,D)$ is a finitely summable regular spectral triple, $D$ is invertible and $A\in\Omega_D^1(\Acal)$ is a self-adjoint gauge potential. Write $\zeta_D(s):=\Tr(|D|^{-s})$. The following holds:
\begin{enumerate}[label=(\roman*)]
\item the function $\zeta_{D+A}$ extends to a meromorphic function with a most simple poles;
\item the function $\zeta_{D+A}$ is regular at $s=0$;
\item the following equality holds
\[
\zeta_{D+A}(0)-\zeta_D(0)= -\ncint\log\big(1+AD^{-1}\big) = \sum_n \frac{(-1)^n}{n}\ncint \big(AD^{-1}\big)^n.
\]
\end{enumerate}
\end{Thm}
\begin{pf}
See \cite[Thm. 2.4]{Connes5}.
\end{pf}

\begin{Rem}
The above use of $\log(1+AD^{-1})$ needs some comments. Namely the operator $1+AD^{-1}$ does not need to be positive, hence $\log(1+AD^{-1})$ is not defined by the functional calculus. But for the logarithm we do have the power series expansion
\[
\log(1+x)=-\sum_{n=1}^\infty \frac{(-1)^n}{n}\,x^n, \qquad \textrm{ for } |x|<1.
\]
Note, that due to summability of the spectral triple, for $n$ large the residue $\ncint (AD^{-1})^n =0$. So the infinite series $-\sum_n \frac{(-1)^n}{n}\,\ncint (AD^{-1})^n$ is in fact a finite sum. This sum is well defined for all gauge potentials $A$. So we use $-\sum_n \frac{(-1)^n}{n}\,\ncint (AD^{-1})^n$ as a definition for $\ncint \log(1+AD^{-1})$. \\
We know that if $D$ is invertible and if $A\in\Omega_D^1(\Acal)$ is a gauge potential, then $A$ is bounded. So there exists a constant $a>0$ such that $\|AD^{-1}\|<a$. For such an $a$ the operator
\[
\log\big(1+\frac{1}{a}AD^{-1}\big)
\]
is well-defined, in which case of course both interpretations (the functional calculus and the finite sum) in Theorem \ref{variation_zeta} are equal.
\end{Rem}

The following proposition relates the residues of the triples $\Scal$ and $\Scal_z$. It can also be used to compute anomalous graphs. For example the result for $n=3$ of the proposition corresponds to the following graph.

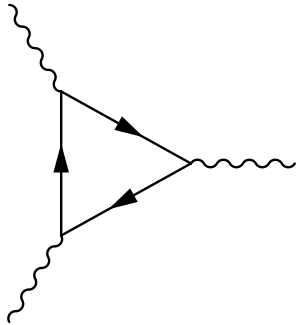
\begin{center}
\begin{fmffile}{graph2}
\begin{fmfgraph}(120,120)
\fmfleft{i1,i2}
\fmfright{o1}
\fmf{fermion,tension=0.4}{v1,v2,v3,v1}
\fmf{photon}{i1,v1}
\fmf{photon}{i2,v2}
\fmf{photon}{o1,v3}
\end{fmfgraph}
\end{fmffile}
\end{center}

\begin{Prop}[Connes \& Marcolli \cite{Connes2}]\label{Connes_1.242}
Suppose the Dirac operator $D$ is invertible, let $A\in \Omega_D^1(\Acal)$ and $n\in\Nat\setminus\{0\}$. Then the function
\[
z\mapsto \tau'\big(((A\otimes1)D_z^{-1})^n\big)
\]
has at most a simple pole at $z=0$, with residue given by
\[
\res_{z=0} \tau'\big(((A\otimes1)D_z^{-1})^n\big) = -\ncint (AD^{-1})^n.
\]
\end{Prop}
\begin{pf}
See \cite[Prop. 1.242]{Connes2}.
\end{pf}

\begin{Thm}\label{fluctuation}
Suppose $(\Acal,\Hcal,D;\gamma)$ is an even $p^+$-summable regular spectral triple, $A\in\Omega_D^1(\Acal)$ is a self-adjoint gauge potential and $D$ is invertible. Then
\begin{equation}\label{eq-fluctuation1}
\zeta_{D+A}(0)-\zeta_D(0)=\sum_n \frac{(-1)^n}{n}\ncint \big(AD^{-1}\big)^n = \sum_n \frac{(-1)^n}{n} \res_{z=0} \tau'\big(((A\otimes1)D_z^{-1})^n\big) .
\end{equation}
\end{Thm}
\begin{pf}
The proof is a simple combination of the results in this section. The first equality of \eqref{eq-fluctuation1} is given by Theorem \ref{variation_zeta} and the second equality follows from Proposition \ref{Connes_1.242}.
\end{pf}

Applying the above theorem to zeta-function regularisation shows that we have two equivalent ways to renormalise the quantity $\Tr\big(\log(1+AD^{-1})\big)$. Namely write $A(s) := \Tr(\log(1+AD^{-1})|D|^{-s})$, then
\begin{align*}
A_{\textrm{ren}} &= \Big(A(s)+\frac{1}{s}\,(\zeta_{D+A}(0)-\zeta_D(0))\Big)|_{s=0}\\
&= \Big(A(s)+\frac{1}{s}\,\sum_{n}\frac{(-1)^n}{n}\,\res_{z=0}\tau'\big(((A\otimes1)D_z^{-1})^n\big)\Big)|_{s=0}.
\end{align*}

\section{Appendix}
In this appendix we will recall some results, most of them related to traces and measures, which are being used in this paper.

Recall that if $T$ is densely defined and symmetric linear operator, then $T^*$ is the closure of $T$. So an equivalent definition of essentially self-adjointness is that $T$ has a unique self-adjoint extension.

\begin{Def}\label{bounded_vector}
Suppose $T:\Dom(T)\ra\Hcal$ is an (unbounded) operator. Denote $\Dom^{\infty}(T):=\bigcap_{n=1}^{\infty}\Dom(T^n)$. If $x\in\Dom^{\infty}(T)$ and there exists a constant $B>0$ (dependent on $x$) such that $\|T^nx\|\leq B^n$ for all $n$, then $x$ is called a {\it bounded vector}. If there exists a constant $C>0$ such that $\|T^nx\|\leq C^n n!$ for all $n$, then $x$ is called an {\it analytic vector}. We denote all bounded vectors of $T$ by $\Dom^b(T)$ and all analytic vectors by $\Dom^a(T)$. It is clear that $\Dom^b(T)$ and $\Dom^a(T)$ are linear subspaces and $\Dom^b(T)\subset\Dom^a(T)$.
\end{Def}

\begin{Prop}\label{Nelson}
Suppose $T$ is an unbounded operator on $\Hcal$ with domain $\Dom(T)$. The following holds:
\begin{enumerate}[label=(\roman*)]
\item if $T$ is self-adjoint, then $\Dom^b(T)\subset \Hcal$ dense;
\item (Nelson's theorem) if $T$ is symmetric and $\Dom^a(T)\subset \Hcal$ is dense, then $T$ is essentially self-adjoint;
\item if $T$ is closed and symmetric, then $T$ is self-adjoint if and only if $\Dom^a(T)\subset\Hcal$ dense.
\end{enumerate}
\end{Prop}
\begin{pf}
Item (i) is \cite[Lemma 7.13]{Schmudgen}. Item (ii) is Nelson's theorem, \cite[Theorem 7.16]{Schmudgen}. Assertion (iii) follows directly from (i) and (ii) with the observation that $\Dom^b(T)\subset\Dom^a(T)$.
\end{pf}

A trace on a von Neumann algebra can be considered as a noncommutative integral. For integrals we have the H\"older inequality and Fubini's theorem. We therefore expect that such results generalise to traces. Such a generalisation is true if one assumes some regularity conditions on the trace. E.g. \cite[Thm. IX.2.13]{Takesaki2} if $\Mcal$ is a semifinite von Neumann algebra with a faithful, semifinite, normal trace $\tau$, for all $x,y\in\Mcal$ it holds that
\begin{equation}\label{holder}
\tau(|xy|)\leq \|x\|\,\tau(|y|).
\end{equation}
The analogue of Fubini is more complicated, we need the following definition.

\begin{Def}
Suppose $(S,\Sigma,\nu)$ is a finite measure space, $\Mcal\subset B(\Hcal)$ a semifinite von Neumann algebra and $\rho$ a semifinite faithful normal trace on $\Mcal$. Denote
\[
\Lcal^1(\Mcal,\rho):= \{T\in\Mcal\,:\, \rho(|T|)<\infty\}.
\]
A bounded function $f:S\ra \Lcal^1(\Mcal,\rho)$ is {\it $*$-measurable} if for all $h\in\Hcal$ the functions $f(\cdot)h:S\ra \Hcal$ and $f(\cdot)^*h:S\ra\Hcal$ are measurable. Define
\[
\Lcal^{so^*}_\infty\big(S,\nu,\Lcal^1(\Mcal,\rho)\big):=\{f:S\ra\Lcal^1(\Mcal,\rho)\,:\, f \textrm{ is } \|\cdot\|\textrm{-bounded },\, *\textrm{-measurable}\}.
\]
\end{Def}

\begin{Prop}\label{interchange_trace_int}{\normalfont\cite[Lemma 3.10]{Azamov}}
Let $\Mcal\subset B(\Hcal)$ be a semifinite von Neumann algebra with semifinite faithful normal trace $\rho$ and $(S,\Sigma,\nu)$ a finite measure space. Assume that the function $f\in\Lcal^{so^*}_\infty(S,\nu;\Lcal^1(\Mcal,\rho))$ and $f$ is uniformly $\Lcal^1(\Mcal,\rho)$-bounded (i.e. there exists $C>0$ such that $\rho(|f(s)|)<C$ for all $s\in S$), then $\int_S f(s)\,d\nu \in\Lcal^1(\Mcal,\rho)$, the function $\rho(f(\,\cdot\,))$ is measurable with respect to the $\sigma$-algebra $\Sigma$ and
\[
\rho\Big(\int_S f(s)\,d\nu(s)\Big) = \int_S \rho(f(s))\,d\nu(s).
\]
\end{Prop}

The following lemma is an immediate consequence of the properties of a spectral measure.
\begin{Lemma}
Let $\Sigma$ be a $\sigma$-algebra on a set $\Omega$. Let $\Ncal$ be a von Neumann algebra and $\tau:\Ncal\ra [0,\infty]$ be a normal trace. Suppose $E:\Sigma\ra B(\Hcal)$ is a spectral measure such that $E(A)\in\Ncal$ for all $A\in\Sigma$. Then
\begin{equation}\label{tau-measure}
\mu_{\tau,E}(A):=\tau(E(A))
\end{equation}
defines a measure on $\Sigma$. If $\tau$ is a finite trace, then $\mu_{\tau,E}$ is a finite measure.
\end{Lemma}

In general from a measure we can construct an integral. In the case the spectral measure is given as the spectral decomposition of a self-adjoint operator we obtain the following link between the functional calculus and integration with respect to the measure $\mu_{\tau,E}$.

\begin{Thm}\label{integration_trace}
Suppose $T$ is a self-adjoint operator on a Hilbert space $\Hcal$ with spectral decomposition $T=\int_{\sigma(T)} \lambda\,dE$. Let $\Ncal\subset\Bcal(\Hcal)$ be a von Neumann algebra, with a normal trace $\tau:\Ncal_{+}\ra[0,\infty]$ and assume that $T$ is affiliated with $\Ncal$. If $f:\sigma(T)\ra\Com$ is a Borel-measurable function such that $f\geq0$ or $f\in\Lcal^1(\sigma(T),\Bcal(\sigma(T)),\mu_{\tau,E})$ then,
\begin{equation}\label{eq-integration_trace1}
\int_{\sigma(T)} f \,d\mu_{\tau,E} = \tau(f(T)).
\end{equation}
\end{Thm}
\begin{pf}
To prove this theorem we will apply the standard machine of measure theory. So first suppose $f=1_A$ for some $A\in\Bcal(\sigma(T))$. Then since $T$ is affiliated with $\Ncal$ each of its spectral projections $E(A)\in\{T\}''\subset\Ncal$. In particular since $E(A)$ is a projection $E(A)\in\Ncal_{+}$. Then we obtain
\[
\int_{\sigma(T)} f\,d\mu_{\tau,E} = \int_{\sigma(T)} 1_A\,d\mu_{\tau,E} = \mu_{\tau,E}(A) = \tau(E(A)) = \tau\Big(\int 1_A\,dE\Big) = \tau(1_A(T)) = \tau(f(T)).
\]
Now for a simple function $f = \sum_{n=1}^{N}\alpha_n A_n$ equation \eqref{eq-integration_trace1} holds, because the integral, the trace and the functional calculus are linear functionals.\\
If $f$ is a positive measurable function, then there exists a sequence of simple functions $(f_n)_n$ with $f_n\uparrow f$ pointwise. The functional calculus implies that
\[
\langle (f_n(T)-f(T))x,y\rangle = \int f_n-f\,dE_{x,y},
\]
which tends to $0$ as $n\ra\infty$ because of the monotone convergence theorem (here $E_{x,y}$ is the measure given by $E_{x,y}(A)=\langle E(A)x,y\rangle$). By construction the operators $f_n(T)\in\Ncal$. Since $\Ncal$ is a von Neumann algebra it is WOT-closed, hence $f(T)\in\Ncal$. Clearly $f(T)\in\Ncal_+$, because $f$ is positive. Using the fact that $\tau$ is a normal trace gives $\tau(f_n(T))\ra\tau(f(T))$. Again an application of the monotone convergence theorem yields
\begin{equation}\label{eq-integration_trace2}
\tau(f(T)) = \lim_{n\ra\infty}\tau(f_n(T)) = \lim_{n\ra\infty}\int_{\sigma(T)}f_n\,d\mu_{\tau,E} = \int_{\sigma(T)}\limsup_{n\ra\infty}f_n\,d\mu_{\tau,E} = \int_{\sigma(T)}f\,d\mu_{\tau,E}.
\end{equation}
This proves the proposition in the case that $f$ is positive. \\
If $f$ is not necessarily positive, but $f\in\Lcal^1(\sigma(T),\Bcal(\sigma(T)),\mu_{\tau,E})$, then $\int_{\sigma(T)}|f|\,d\mu_{\tau,E}<\infty$. We split up $f$ in four parts, $f=\re(f)^{+}-\re(f)^{-}+i\im(f)^{+}-i\im(f)^{-}$ and apply the previous equality \eqref{eq-integration_trace2} to each of the four summands. It follows that $\int_{\sigma(T)}\re(f)^+\,d\mu_{\tau,E}<\infty$ and similar statements for $\re(f)^{-},\im(f)^{+},\im(f)^{-}$. By linearity of the integral and trace we obtain $\int_{\sigma(T)} f \,d\mu_{\tau,E} = \tau(f(T))$, as desired.
\end{pf}

Note that for a projection $P:\Hcal\ra\Hcal$, $\Tr(P)=\dim(\ran(P))$. So the following definition gives a natural generalisation of compact operators.

\begin{Def}
Let $\Mcal$ be a von Neumann algebra acting on $\Hcal$ and $\tau$ be a trace on $\Mcal$. We denote by $\Pcal(\Mcal):=\{P\in\Mcal\,:\, P \textrm{ is a projection}\}$, the projections. The $\textrm{span}(\{P\in\Pcal(\Mcal)\,:\, \tau(P)<\infty\})$, are called the {\it $\tau$-finite rank operators}. The {\it $\tau$-compact operators} are defined as $\Kcal(\Mcal,\tau):= \clo(\Rcal(\Mcal,\tau))$, where the closure is taken in the norm topology.
\end{Def}

It is immediate that if $\Mcal$ is a von Neumann algebra acting on $\Hcal$ and $\tau$ a trace on $\Mcal$, then $\Kcal(\Mcal,\tau)$ is a closed ideal in $B(\Hcal)$ in the norm-topology. Furthermore $\Kcal(\Mcal,\tau)$ equals the closed ideal generated by the $\tau$-finite projections.

\begin{Thm}\label{thm-tau-discrete}
Let $\Mcal$ be an infinite, semifinite von Neumann algebra acting on $\Hcal$, equipped with a normal, faithful trace $\tau$. Suppose $T$ is a self-adjoint $\Mcal$-affiliated  (unbounded) operator on $\Hcal$ with spectral decomposition $T=\int \lambda\,dE$. Then the following are equivalent:
\begin{enumerate}[label=(\roman*)]
\item for all $\lambda\notin \sigma(T)$ it holds $(T-\lambda)^{-1}\in \Kcal(\Mcal,\tau)$;
\item there exists a $\lambda_0\notin \sigma(T)$ such that $(T-\lambda_0)\in\Kcal(\Mcal,\tau)$;
\item for all $\lambda\in\Rea,\; \tau(E([-\lambda,\lambda]))<\infty$.
\end{enumerate}
\end{Thm}
\begin{pf}
Clearly (i) implies (ii). The converse implication is immediate from the first resolvent formula \cite[Thm. VIII.2]{Reed1} and the fact that $\Kcal(\Mcal,\tau)$ is an ideal in $B(\Hcal)$.\\
Now suppose (iii) holds.  Let $\lambda_0\notin\sigma(T)$. Define the bounded measurable function
\[
f:\sigma(T)\ra\Com,\qquad \lambda\mapsto\,\frac{1}{\lambda-\lambda_0}.
\]
Then $f(T)=(T-\lambda_0)^{-1}$ is a bounded operator. Put $f_n:=f1_{[-n,n]}$ and define $S_n:=f_n(T)$. Since $f_n$ is bounded and measurable and the operator $T$ is affiliated to $\Mcal$ we have $S_n\in\Mcal$. Let $Q_n$ be the orthogonal projection on $S_n(\Hcal)$. Then $Q_n\leq E([-n,n])$, thus by assumption $\tau(Q_n)\leq \tau(E([-n,n]))<\infty$. So the operators $S_n$ have $\tau$-finite rank. Now let $\eps>0$. Select $N\in\Nat$ such that $|\frac{1}{N-\lambda_0}|,\, |\frac{1}{-N-\lambda_0}|<\eps$. For $n>N$ and $x\in\Hcal$, we have
\begin{align}\label{compact-resolvent2}
\|S_nx - (T-\lambda_0)^{-1}x\| &= \Big\|\big(\int_{[-n,n]}\frac{1}{\lambda-\lambda_0}\,dE\big)\,x - \big(\int_{\Rea}\frac{1}{\lambda-\lambda_0}\,dE\big)\,x\Big\|\notag\\
&= \Big\|\big(\int_{(-\infty,-n)\cup(n,\infty)}\frac{1}{\lambda-\lambda_0}\,dE\big)\,x\Big\|\notag\\
&\leq \eps\|x\|.
\end{align}
This establishes (i).\\
Now suppose (i) holds we will show (iii) is true. Because $\sigma(T)\subset\Rea$, the operator $(T-i)^{-1}$ is $\tau$-compact. So there exists a sequence of $\tau$-finite rank operators $(S_n)_n$ such that $\|S_n-(T-i)^{-1}\|\ra0$. Since $(T-i)^{-1} = \int_{\Rea} (\lambda-i)^{-1}\,dE$, for all $\eps>0$ there exist $N\in\Nat$ and $\nu_0>0$ such that for all $n>N$ and $\nu>\nu_0$ it holds that
\begin{equation}\label{compact-resolvent1}
\Big\|S_n-\int_{[-\nu,\nu]} \frac{1}{\lambda-i}\,dE\Big\|\leq \big\|S_n - (T-i)^{-1}\big\| + \Big\| (T-i)^{-1} - \int_{[-\nu,\nu]} \frac{1}{\lambda-i}\,dE\Big\| \leq\eps/2+\eps/2 = \eps.
\end{equation}
Now suppose that $\tau(E([-\mu_0,\mu_0]))=\infty$. Let $\eps<\frac{1}{\mu_0^2+1}$ and select corresponding $N$ and $\nu_0$. Denote by $Q_n$ again the projection on the range of $S_n$. By orthogonality of $Q_n$ and $1-Q_n$ we obtain
\begin{align*}
\tau(E([-\mu_0,\mu_0])) &= \tau(E([-\mu_0,\mu_0]\wedge Q_n + E([-\mu_0,\mu_0])\wedge (1-Q_n))\\
&\leq \tau(Q_n) + \tau(E([-\mu_0,\mu_0])\wedge (1-Q_n)).
\end{align*}
Since $\tau(Q_n)<\infty$ for all $n$
\[
\tau(E([-\mu_0,\mu_0])\wedge (1-Q_n))\geq \tau(E([-\mu_0,\mu_0])) - \tau(Q_n) = \infty.
\]
Hence $(E([-\mu_0,\mu_0])\wedge (1-Q_n))\Hcal\neq\{0\}$. Select an element $x\in E([-\mu_0,\mu_0])\Hcal\cap\ker(S_n)$, with $\|x\|=1$. For this $x$ and for any $\nu>\mu_0$ the following identities hold
\[
\Big(\int_{[-\nu,\nu]}\frac{1}{\lambda-i}\,dE\Big) x = \Big(\int_{[-\mu_0,\mu_0]}\frac{1}{\lambda-i}\,dE\Big) x, \qquad S_nx=0.
\]
Combining these equalities and using the properties of $E$ gives
\begin{align*}
\Big\|S_n x-\big(\int_{[-\nu,\nu]} \frac{1}{\lambda-i}\,dE\big)\, x\Big\|
&= \Big\|\big(\int_{[-\mu_0,\mu_0]} \frac{1}{\lambda-i}\,dE\big)\, x\Big\|\\
&\geq \Big|\Big\langle\big(\int_{[-\mu_0,\mu_0]} \frac{\lambda+i}{\lambda^2+1}\,dE\big)\, x,x\Big\rangle\Big|\\
&\geq \Big|\im\big(\int_{[-\mu_0,\mu_0]} \frac{\lambda+i}{\lambda^2+1}\,dE_{x,x}\big)\Big|\\
&=\Big|\int_{[-\mu_0,\mu_0]} \frac{1}{\lambda^2+1}\,dE_{x,x}\Big|\\
&>\eps\, |E_{x,x}([-\mu_0,\mu_0])|= \eps.
\end{align*}
But this is a contradiction with equation \eqref{compact-resolvent1}. Hence we established (iii).
\end{pf}

\begin{Def}\label{tau_compact}
If $\Mcal$ is a semifinite von Neumann algebra acting on $\Hcal$, with a normal, faithful trace $\tau$ and if $T$ is a self-adjoint $\Mcal$-affiliated operator on $\Hcal$ which satisfies one of the equivalent conditions of Theorem \ref{thm-tau-discrete}, then $T$ is called a {\it $\tau$-discrete operator}.
\end{Def}

\end{document}